\documentclass[%
 reprint,
amsmath,amssymb,
prl,
]{revtex4-1}
\usepackage{graphicx}
\usepackage{dcolumn}
\usepackage{bm}
\usepackage{xcolor}
\usepackage[normalem]{ulem}

\usepackage{pdfpages}

\makeatletter
\AtBeginDocument{\let\LS@rot\@undefined}
\makeatother
\begin{document}

\title{Anomalous Coulomb Drag between InAs Nanowire and Graphene Heterostructures}

\author{Richa Mitra$^1$, Manas Ranjan Sahu$^1$, Kenji Watanabe$^2$, Takashi Taniguchi$^2$, Hadas Shtrikman$^3$, A.K Sood$^1$ and Anindya Das$^1$
\footnote{anindya@iisc.ac.in}
}


\affiliation{ $^1$Department of Physics,Indian Institute of Science, Bangalore, 560012, India.\\
$^2$National Institute for Materials Science, Namiki 1-1, Ibaraki 305-0044, Japan. \\
$^3$Department of Physics, Weizmann Institute of Technology, Israel.}

\date{\today}

\begin{abstract}
Correlated charge inhomogeneity breaks the electron-hole symmetry in two-dimensional (2D) bilayer heterostructures which is responsible for non-zero drag appearing at the charge neutrality point. Here we report Coulomb drag in novel drag systems consisting of a two-dimensional graphene and a one dimensional (1D) InAs nanowire (NW) heterostructure exhibiting distinct results from 2D-2D heterostructures. For monolayer graphene (MLG)-NW heterostructures, we observe an unconventional drag resistance peak near the Dirac point due to the correlated inter-layer charge puddles. The drag signal decreases monotonically with temperature ($\sim T^{-2}$) and with the carrier density of NW ($\sim n_{N}^{-4}$), but increases rapidly with magnetic field ($\sim B^{2}$). These anomalous responses, together with the mismatched thermal conductivities of graphene and NWs, establish the energy drag as the responsible mechanism of Coulomb drag in MLG-NW devices. In contrast, for bilayer graphene (BLG)-NW devices the drag resistance reverses sign across the Dirac point and the magnitude of the drag signal decreases with the carrier density of the NW ($\sim n_{N}^{-1.5}$), consistent with the momentum drag but remains almost constant with magnetic field and temperature. This deviation from the expected $T^2$ arises due to the shift of the drag maximum on graphene carrier density. We also show that the Onsager reciprocity relation is observed for the BLG-NW devices but not for the MLG-NW devices. These Coulomb drag measurements in dimensionally mismatched (2D-1D) systems, hitherto not reported, will pave the future realization of correlated condensate states in novel systems.
\end{abstract}

\maketitle

Correlated electronic states continue to be the focus of the condensed matter community, thanks to their rich complexity in physics and fascinating technological potential in the near future. Over the years the search for realizing highly correlated states have led to the discovery of novel many-body states like excitonic condensate states  \cite{kellogg2002observation,seamons2009coulomb,li2017excitonic,nandi2012exciton}, fractional quantum Hall states \cite{eisenstein1992new,suen1992observation}, Luttinger liquid phase\cite{laroche20141d,debray2001experimental,yamamoto2006negative,flensberg1998coulomb}  etc. Coulomb drag has proven to be the quintessential tool for probing the electron-electron interaction in correlated systems and studied in diverse set of systems like 2D electron gas (2DEG) based (AlGaAs/GaAs) heterostructures \cite{solomon1989new,gramila1991mutual,kellogg2003bilayer,kellogg2002observation,seamons2009coulomb,nandi2012exciton,croxall2008anomalous} to quantum wires \cite{laroche20141d,debray2001experimental,yamamoto2006negative,flensberg1998coulomb}
. 
In Coulomb drag, current ($I_{D}$) passing in one of the layers produces an open circuit voltage ($V_{D}$) in the other layer without any particle exchange. Very recently, graphene based heterostructures \cite{gorbachev2012strong,kim2011coulomb,li2016negative,lee2016giant,gamucci2014anomalous} have revealed intriguing feature of the drag signal at the Dirac point \cite{gorbachev2012strong,li2016negative,lee2016giant}; namely that it can have both positive \cite{gorbachev2012strong} and negative \cite{lee2016giant} amplitudes. A puzzling feature is its temperature dependence which shows monotonic behavior with a maximum at the lowest temperature in BLG \cite{lee2016giant} whereas non-monotonic variation with a maximum at an intermediate temperature ($\sim$ 100K) for MLG \cite{gorbachev2012strong}. The drag signal at the Dirac point can not be explained by the conventional momentum drag mechanism involving the momentum transfers from the drive to the drag layer, and hence two new mechanisms; namely Energy drag \cite{song2012energy,song2013coulomb,song2013hall} and inhomogeneous momentum drag \cite{ho2018theory} have been proposed. 


\begin{figure*}
\includegraphics[width=1\textwidth]{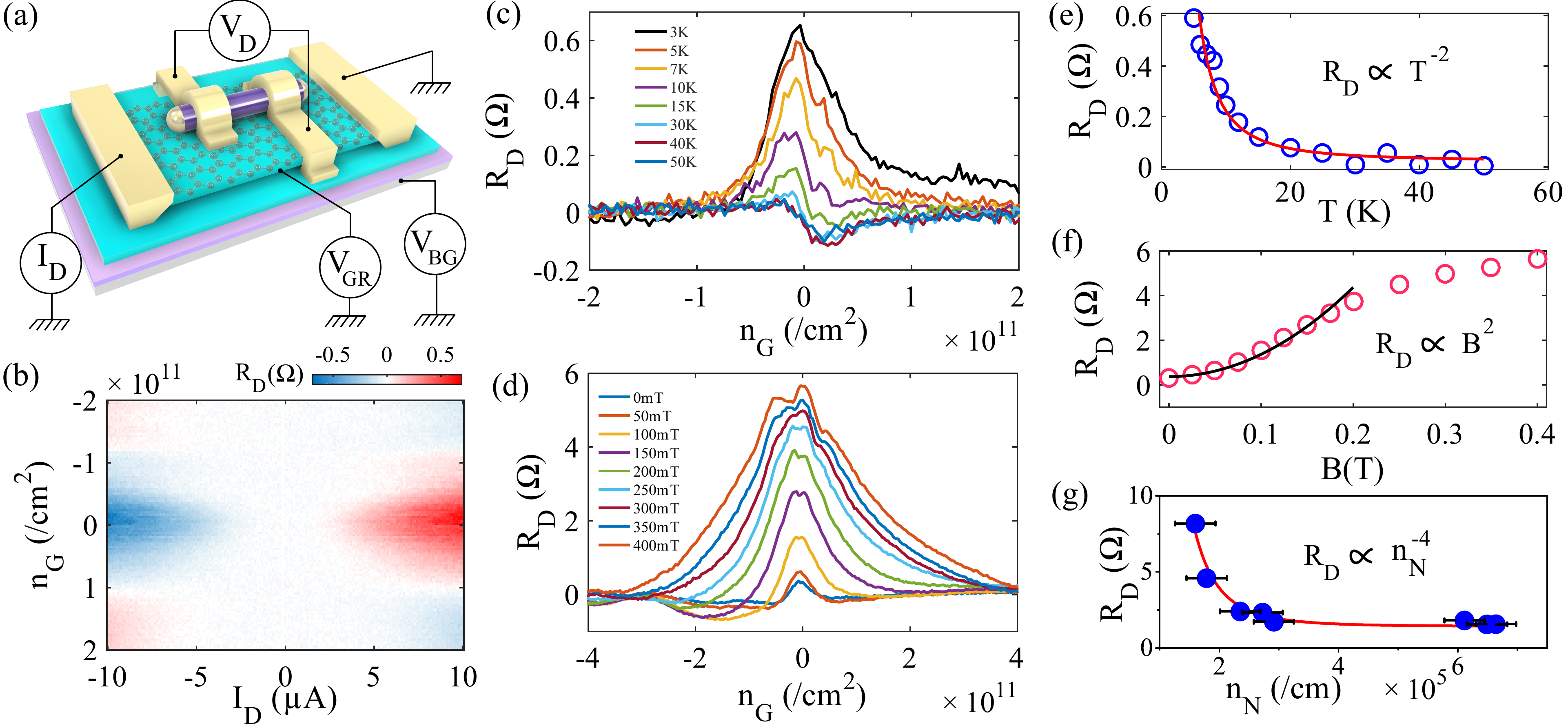}
\caption{\label{fig:epsart}  (a) Device schematic : The heterostructure consists of a InAs nanowire on top of a  hBN encapsulated graphene stack assembled on $Si/SiO_{2}$ substrate. (b) 2D colormap of $R_{D}$ at $T$=1.5K plotted against the $I_{D}$ and $n_{G}$ for the D1 device (MLG-NW). (c) Response of the $R_{D}$ at different temperatures. (d) $R_{D}$ versus $n_{G}$ plot for different magnetic fields at $T$=1.5K. (e) Peak values of the $R_{D}$ (blue circles) plotted with temperature. $R_{D}$ decreases with temperature and fits (red solid line) well with $R_{D} \propto T^{-2}$. (f) The pink open circles are the peak value of $R_{D}$ plotted with magnetic field at $T$=1.5K. The black solid line shows data upto 200mT fits well with $B^2$. (g) $R_{D}$ at the Dirac point as a function of $n_{N}$ at $T$=1.5K. The solid line is an overlay of $n_{N}^{-4}$ with the data. The errorbars in $n_{N}$ have been estimated from different sweeps of measurements of device shown in Fig S-3B (a) of the Supplemental Material  \cite{supplement}.}
\end{figure*}

A new drag system consisting of 2D graphene and a confined 1D nanowire or nanotube, not only has a potential for probing the graphene locally, but also the dimensionally mismatched Coulomb drag system can potentially become the foreground for studying the effect of dimension on scattering mechanisms in Coulomb drag \cite{sirenko1992influence, lyo2003coulomb}. This kind of drag system is expected to show novel quantum phases in the strong coupling regime \cite{narozhny2016coulomb} in addition to being a tool for studying the graphene hydrodynamics near the Dirac point \cite{lucas2018hydrodynamics}. With this motivation we have carried out the Coulomb drag experiments in MLG-InAs NW and BLG-InAs NW devices as a function of density ($n$), temperature ($T$) and magnetic field ($B$). The MLG-NW devices show a drag resistance ($R_{D} = V_{D}/I_{D}$) maximum around the Dirac point and its dependence on $n$, $T$ and $B$ establish the Energy drag as the dominant mechanism. In comparison, absence of the drag signal at the Dirac point for the BLG-NW devices and flipping sign across the Dirac point with negligible dependence on $T$ and $B$ suggest the dominance of momentum drag mechanism.
  
The device and measurement configuration are schematically presented in Fig. 1a. All the devices comprise of heterostructures of hexagonal boron nitride (hBN) encapsulated graphene stack and InAs NW with diameter between 50 to 70 nm. The heterostructures were assembled by the standard hot pick up technique \cite{wang2013one,pizzocchero2016hot}, where the $\sim$ 10 nm thick top hBN of the graphene stack separates the graphene channel and the NW (SI-1 of Supplemental Material  \cite{supplement}). The inhomogeneity ($\delta n$) of graphene is $\sim 2.5 \times 10^{10} /cm^{2}$, which corresponds to a Fermi energy broadening of $\Delta_{0} \sim 15 meV$ and $ \sim 0.5 meV$ for MLG and BLG, respectively. The NWs could only be electron doped due to Fermi energy pinning near the conduction band.  The 1D nature of the NW used is ascertained by measuring the electrical conductance as a function of the $V_{BG}$ for shorter channel length showing participation of 3-5 sub-bands (see SI-1E of Supplemental Material \cite{supplement}). The charge inhomogeneity in the NW was investigated by measuring the temperature-dependent conductance as shown in Fig. S-1F of Supplemental Material  \cite{supplement}, which suggests the localization length of $\sim$ 100-200 nm. All the measurements were done in a He4 cryostat in the temperature range of 1.5K to 200K. 

\begin{figure*}
\includegraphics[width=1\textwidth]{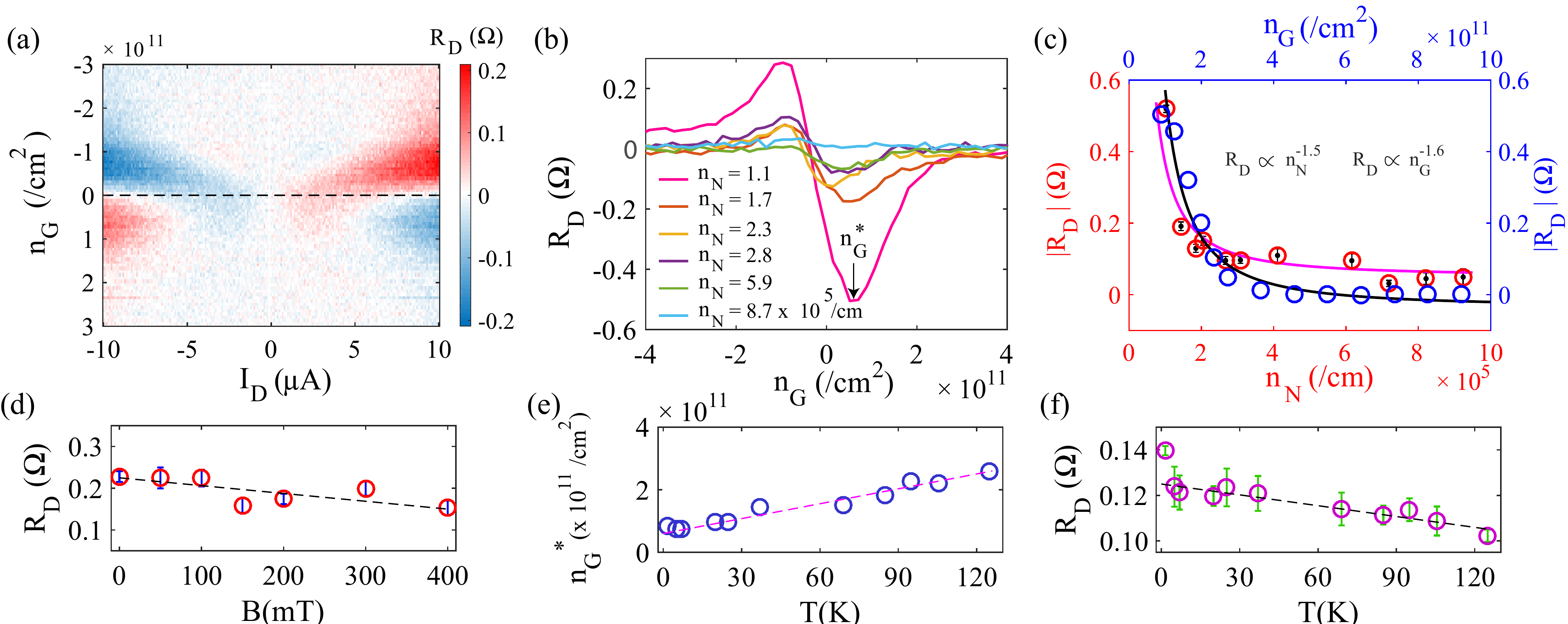}
\caption{\label{fig:epsart} (a) 2D colormap of $R_{D}$ with $I_{D}$ and $n_{G}$ at $T$=1.5K, $V_{GR}=1V$ for a BLG-NW device. The horizontal dashed line is the Dirac point of the graphene. (b) $R_{D}$ versus $n_{G}$ plot at $T$=1.5K for different $n_{N}$ tuned by the $V_{GR}$ from 0.9 to 5V. (c) The red circles are the plot for dip value of $R_{D}$ at different $n_{N}$. 
The variation of the drag signal with $n_{G}$ at $V_{GR}=0.9V$ is indicated by the blue open circles. The solid lines are the fitting to $\sim n_{N}^{-1.5}$ and $n_{G}^{-1.6}$. (d) The variation of $R_{D}$ with magnetic field at $T$=1.5K. (e) The position of the dip ($n_{G}^{*}$) of $R_{D}$ as a function of temperature (raw data in Fig. S-4B of Supplemental Material \cite{supplement}). (f) The dip value of $R_{D}$ plotted as a function of temperature. The dashed lines in d, e and f are the guiding lines.}
\end{figure*}  
    
The drag measurements were performed by the DC technique, where $I_{D}$ was passed through the graphene and $V_{D}$ was measured on the NW as shown in Fig. 1a or vice versa. The carrier density of the graphene ($n_{G}$) and NWs ($n_{N}$) were tuned by the $SiO_{2}$ back gate ($V_{BG}$) and by a voltage ($V_{GR}$) between the graphene and the NW (SI-2 of Supplemental Material \cite{supplement}). In our DC measurements, the drag signal contains a predominant flipping component (sign reversal of the drag voltage with  $I_{D}$) together with a small non-flipping component. 
Here, we present the extracted flipping part (in the linear regime) as mentioned in section SI-2B of Supplemental Material  \cite{supplement}, which is consistent with the drag signal measured by the low-frequency AC (at 7Hz) technique (SI-2 of Supplemental Material \cite{supplement}). The tunneling resistance of the $\sim$ 10nm thick hBN between the graphene and NWs was more than $5-10 G\Omega$ in all the devices. We have used two MLG-NW (D1, D2) and two BLG-NW (D3, D4) devices for the drag measurements.

Figure 1b shows the 2D colormap for the MLG-NW device (D1) at $T$=1.5K and $n_{N} \sim 4 \times 10^{5}$ cm$^{-1}$, where $R_{D}$ is plotted with $I_{D}$ varying from -10$\mu A$ to +10$\mu A$ and $n_{G}$ varying from $0$ to $2 \times 10^{11} /cm^2$ for both electron and hole doping. 
The drag signal peaks near the Dirac point and subsequently decays at higher $n_{G}$. Figure 1c shows $R_{D}$ at different temperatures. The peak magnitude decreases rapidly with temperature as shown by open circles in Fig. 1e. 
Figure 1d shows the dependence of $R_{D}$ on magnetic field upto 400 mT at $T$=1.5K. Notably, we observe a giant increase (by one order of magnitude) of the drag peak with increasing magnetic field as shown by open circles in Fig. 1f. The dependence of the drag peak on $n_{N}$ measured by varying $V_{GR}$ (SI-3C of Supplemental Material \cite{supplement}) is shown in Fig. 1g.

Figure 2a shows the 2D colormap for the BLG-NW device (D3), where $R_{D}$ is plotted as a function of $I_{D}$ and $n_{G}$ at $T$=1.5K for $n_{N} \sim 1.3 \times 10^{5}$ cm$^{-1}$. 
In contrast to MLG-NW devices, the drag signal flips sign from positive to negative as $n_{G}$ shifts from holes to electrons across the Dirac point with distinct peak and dip at finite densities of holes and electrons. At the Dirac point the $R_{D}$ is negligible unlike the MLG-NW device. 
Figure 2b shows $R_{D}$ as a function of $n_{G}$ for different NW densities ($n_{N} \sim 1$ to $10 \times 10^{5}$ cm$^{-1}$) tuned by $V_{GR}$. 
The blue circles in Fig. 2c quantify how the magnitude of $R_{D}$ decreases with $n_{G}$ (for electron side in Fig. 2b) for $n_{N} = 1.1 \times 10^{5}$ cm$^{-1}$, whereas the red circles show the magnitude of the dip of $R_{D}$ at $n_{G}^{*}$ (marked in Fig. 2b) as a function of $n_{N}$. 
Figure 2d shows that $R_{D}$ at $n_{G}^{*}$ for the BLG-NW device remains almost constant with magnetic field (raw data in SI-4A of Supplemental Material \cite{supplement}), in contrast with the MLG-NW device. Figure 2e and 2f demonstrate the temperature dependence of the drag signal for the BLG-NW device. It can be seen from Supplemental Material \cite{supplement} Fig. S-4B (raw data) that the peak (hole side) or dip (electron side) position of $R_{D}$ shifts towards higher carrier density in graphene with increasing temperature for a fixed carrier density of the NW ($n_{N} \sim 1 \times 10^{5}$ cm$^{-1}$). Figure 2e shows the position ($n_{G}^{*}$) and the corresponding value of $R_{D}$ in Fig. 2f as a function of temperature. Unlike the MLG-NW device, the drag signal in the BLG-NW device clearly displays much less variation with temperature. 

The observations of monotonic decrease of drag signal of the MLG-NW device as well as weak dependence of the drag signal of the BLG-NW device on increasing temperature are anomalous as compared to the conventional momentum drag which predicts $T^2$  \cite{tse2007theory,narozhny2012coulomb,katsnelson2011coulomb,peres2011coulomb}     increase as seen in double-layer MLG heterostructures \cite{gorbachev2012strong}. 
Anomaly in temperature-dependence, specifically, drag signal increasing with lowering temperature has been observed in 2DEG (GaAs-AlGaAs) \cite{solomon1989new} or 2DEG-graphene \cite{gamucci2014anomalous} heterostructures. The anomalous upturn of the drag signal with lowering temperature at low temperature regime indicated the presence of interlayer excitonic condensation in 2DEG-graphene system \cite{gamucci2014anomalous} or the Luttinger liquid state in quantum wire systems \cite{laroche20141d}. The possibility of excitonic condensation in our MLG-NW devices is ruled out as the drag peak appears at the Dirac point of graphene with the NW having a finite density. 

To explain our results, we first recall the three main mechanisms of the Coulomb drag: (i) homogeneous momentum drag (HMD) - momentum transfers via Coulomb mediated scattering, (ii) inhomogeneous momentum drag (IMD) - momentum transfer in presence of correlated inter-layer charge puddles and (iii) energy drag (ED) - vertical energy transfer in presence of correlated inter-layer charge puddles. 
Since the HMD signal should be zero at the Dirac point and increases as $T^2$ in a Fermi liquid \cite{tse2007theory,narozhny2012coulomb,katsnelson2011coulomb,peres2011coulomb}, it can be ruled out as the possible mechanism for our MLG-NW devices. Now, both IMD and Energy drag mechanisms predict a maximum of $R_{D}$ at the Dirac point due to the presence of correlated inter-layer charge puddles, although the underlying physics is different. The effective momentum theory (EMT) of IMD \cite{ho2018theory} suggests an increase of the drag signal with temperature in low temperature regime and should decrease when $k_{B}T > 0.5 \Delta_{0}$. Further, the EMT does not explain the effect of magnetic field on the drag signal. Thus the anomalous decrease with temperature and enhancement of $R_{D}$ with magnetic field in our MLG-NW devices is not consistent with the predictions of the IMD.

Coming now to the Energy drag mechanism, a positive correlation of charge inhomogeneities in MLG and NW gives rise to a positive drag peak around the Dirac point due to the combined effect of Coulomb mediated vertical energy transfer and thermoelectric Peltier effect \cite{song2012energy}. The Energy drag is expected to increase \cite{song2013hall} with magnetic field as $B^{2}$ and display a non-monotonic behavior with temeprature \cite{song2012energy}. Fig. 1f for the MLG-NW device clearly shows $B^2$ dependence of $R_{D}$ at lower magnetic field which is consistent with the Energy drag mechanism \cite{song2012energy}. To explain the temperature dependence, a quantitative theory of ED in 2D-1D system is required. In the absence of such theory, we appeal to Song et.al for 2D-2D system which shows  \cite{song2012energy}  $R_{D} \propto \frac{1}{2T\kappa} \frac{\partial Q}{\partial \mu_{G}} \frac{\partial Q}{\partial \mu_{N}}$, where $\frac{\partial Q}{\partial \mu}$ is the partial derivatives of the Peltier coefficient $Q$ with respect to the chemical potentials of drive ($\mu_{G}$) and drag layers ($\mu_{N}$). The quantity $\kappa$ is the sum of the thermal conductivities ($\kappa_G + \kappa_N$) of the two layers. For double-layer graphene heterostructures, the Energy drag mechanism \cite{song2012energy} generates a non-monotonic temperature behavior where the drag signal increases as $T^2$ upto a temperature equivalent to $\sim$ $\Delta_{0}$ and subsequently decreases as $T^{-4}$. For the MLG-NW devices, the typical value of $\Delta_{0}$ is $\sim 15meV$ (equivalent to 150K). Hence, according to the Energy drag mechanism, the drag signal should have increased monotonically upto $\sim$ 150K. While discussing the Energy drag mechanism, we need to keep in mind that the studies so far assume two identical layers of graphene having similar properties such as mobility, thermal conductivity, electrical conductivity etc. In contrast, we measure the Coulomb drag between two very dissimilar systems: a high mobility 2D graphene sheet and a low-mobility semiconducting 1D NW, with very different electrical transports. More importantly, the NWs have electrical conductivity $ \sim$ 1$e^2/h$ and thus poor electronic thermal conductivity ($\kappa_{e}$) as compared to graphene, making phonon contributions to the thermal transport ($\kappa_{ph}$) dominant \cite{xiong2017significantly}. Hence, $\kappa = \kappa_G + \kappa_N = \kappa_{e} + \kappa_{ph}$. Since $\kappa_{e} \propto T$ and $\kappa_{ph}$ (electron-phonon contribution) $\propto T^5$ (\cite{wellstood1994hot,Srivastaveaaw5798}) , $\kappa$ = $(aT+bT^5)$, where $a$ and $b$ are the relative contributions from the electronic and the phononic parts. The contribution of the interlayer dielectric hBN to $\kappa_{ph}$ is expected to be much smaller than that of the NW and hence is not expected to affect the temperature dependence. Using $\frac{\partial Q}{\partial \mu_{G}}$ $\propto$ $\frac{T^2}{\Delta_{0}^2}$ at the Dirac point and $\frac{\partial Q}{\partial \mu_{N}}$ $\propto$ $\frac{T^2}{\mu_{N}^2}$ at $\mu_{N} \neq 0$ (SI-5 of Supplemental Material \cite{supplement} for details), the temperature dependence of $R_{D}$ is $R_{D} \propto \frac{T^3}{{\mu_{N}^2}{\Delta_{0}}^2(aT+bT^5)}$. Noticeably, the $R_D$ still has the non-monotonic dependence on temperature, depending on the relative magnitudes of the parameters $a$ and $b$. The calculated $R_D$ for different values of $a/b$ is shown in Fig. 3a, where one can see that the crossover happens at temperatures near $\sim 1K$ (below our temperature range) and decreases as $T^{-2}$ consistent with our experimental data (the solid lines in Fig. 1e for D1, and in the inset of Fig. 3a for D2). Furthermore, Fig. 3b shows the similarities between the dependence of $R_{D}$ and $\partial Q/\partial \mu_{G}$ on $n_{G}$ (SI-5 of Supplemental Material \cite{supplement}), which further strengthens the Energy drag to be the dominant mechanism in MLG-NW devices. Moreover, the effect of carrier density of the NW on the drag peak  showing $n_{N}^{-4}$ dependence (Fig. 1g) is compatible with the Energy drag mechanism as the $\frac{\partial Q}{\partial \mu_{N}} \propto \frac{T^2}{\mu_{N}^2} = \frac{T^2}{n_{N}^4}$ (SI-5 of Supplemental Material \cite{supplement}) .

\begin{figure}
\includegraphics[width=0.5\textwidth]{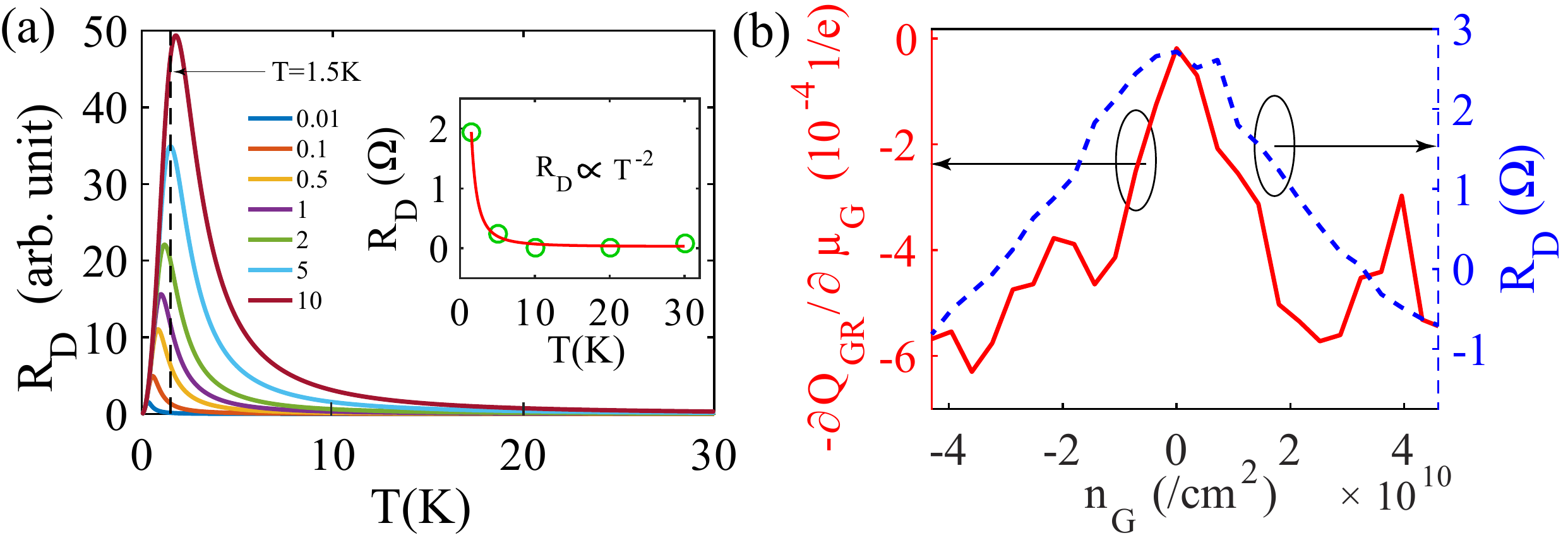}
\caption{\label{fig:epsart} (a) Theoretically calculated $R_{D}$ as a function of temperature for several values of $a/b$ ranging from 0.01 to 10. The inset shows the experimental $R_{D}$ (open circles) as a function of temperature for D2 device with $T^{-2}$ fitting (solid red line). (b) Similarities between $\frac{\partial Q}{\partial \mu}$ (red solid line) and $R_{D}$ (blue dashed line).}
\end{figure} 

We will now discuss the possible drag mechanism for the BLG-NW devices. Drag being almost zero near the Dirac point (Fig. 2a and 2b) rules out Energy drag and IMD, in favour of HMD as a possible mechanism, where $R_{D} \propto \frac{(k_{B}T)^2}{n_{G}^{1.5} n_{N}^{1.5}}$ is consistent with our result as shown in Fig. 2c (solid lines). However, we do not observe the predicted $T^2$ increase of the drag signal (Fig. 2f). This can be due to that the drag signal not only slowly varies with increasing temperature but also the shift of the $R_{D}$ maximum and minimum position ($n_{G}^{*}$) towards higher $n_{G}$ (Fig. 2e and 2f). This happens due to the temperature-induced Fermi energy broadening, over and above the intrinsic disorder limited $\Delta_{0}$ ($\sim$ 0.5 meV in BLG). 

In order to see the effect of dimensionality mismatch on Onsager reciprocity relation, we have measured the drag signal in both NW and graphene as shown in Fig 4. As can be seen the Onsager relation is valid in the BLG-NW device (Fig. 4b), whereas it is violated for the MLG-NW device (Fig. 4a). The violation of Onsager relation has been reported in bilayer 2DEG and 2DEG-graphene devices \cite{croxall2008anomalous,seamons2009coulomb,gamucci2014anomalous}, but the exact reason is not clearly understood. We suggest that the role of different drag mechanisms in Onsager relation can be at play in the dimensionality mismatched devices.

In conclusion, we have performed drag measurements on dimensionally mismatched MLG/BLG-InAs NW heterostructures hitherto not reported. We observe very different drag signals for the MLG-NW and the BLG-NW devices. The MLG-NW devices show a maximum of $R_{D}$ at the Dirac point and the peak value decreases with increasing temperature as well as with the carrier density of the NW. Further, the drag increases by one order of magnitude with magnetic field. These results show that the Energy drag mechanism is dominant for the Coulomb drag in the MLG-NW heterostructures, where the phononic thermal conductivity of the NWs plays a significant role in reduced drag signal with increasing temperature. In contrast, for the BLG-NW devices, the drag reverses sign across the Dirac point as expected from the momentum drag mechanism, with slow variation with temperature and magnetic field. Our results are promising for realizing the correlated states in dimensionally mismatched novel devices, with different mechanisms at play.

\begin{figure}
\includegraphics[width=0.5\textwidth]{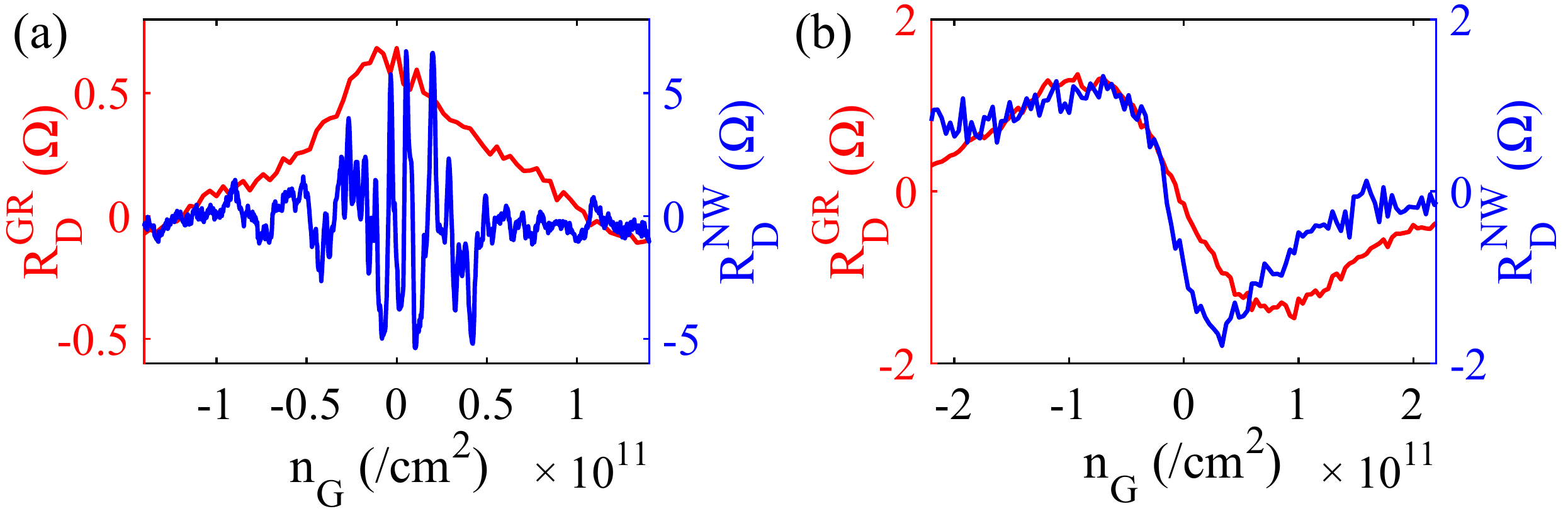}
\caption{\label{fig:epsart} (a) Onsager in MLG-NW device at $T$=1.5K. The red line corresponds to the $R_{D}$ measured on NW whereas the blue line corresponds to the $R_{D}$ measured on MLG. (b) Similar data for the BLG-NW device.}
\end{figure} 

Authors thank Dr. Derek Ho and Prof. Shaffique Adam for useful discussions on in-homogeneous momentum drag. AD thanks DST (DSTO-2051, DSTO-1597) and AKS thanks Year of Science Fellowship from DST for the financial support. K.W. and T.T. acknowledge support from the Elemental Strategy Initiative conducted by the MEXT, Japan and the CREST (JPMJCR15F3), JST. We acknowledge Michael Fourmansky for his professional assistance in NWs MBE growth. HS acknowledges partial funding by Israeli Science Foundation (grant No. 532/12 and grant No. 3-6799), BSF grant No. 2014098 and IMOS-Tashtiot grant No. 0321-4801. HS is an incumbent of the Henry and Gertrude F. Rothschild Research Fellow Chair.

\bibliography{references}

\onecolumngrid
\newpage
\thispagestyle{empty}
\mbox{}


\section*{Supplementary material}

# Supplementary Information: Anomalous Coulomb Drag between InAs Nanowire and Graphene Heterostructures

Richa Mitra[1], Manas Ranjan Sahu[1], Kenji Watanabe[2], Takashi Taniguchi[2], Hadas Shtrikman[3], A.K Sood[1] and Anindya Das[1] *

[1]Department of Physics, Indian Institute of Science, Bangalore, 560012, India.
[2]National Institute for Materials Science, Namiki 1-1, Ibaraki 305-0044, Japan.
[3]Department of Physics, Weizmann Institute of Technology, Israel.

## SI-1. Coulomb drag devices

In this section we furnish all the details about the Coulomb drag devices. The section has been divided into several sub-sections as mentioned below.

### SI-1A. Monolayer Graphene (MLG) /Bilayer Graphene (BLG) -InAs Nanowire Heterostructures

To prepare the hBN-graphene-hBN stacks, at first the graphene and hexagonal boron nitride (hBN) are exfoliated from their respective crystals using the standard scotch tape method [1] and transferred on top of $Si$ substrate caped with thermally grown $SiO_2$ (thickness $\sim$ 300 nm). The monolayer and bilayer graphene flakes are initially identified using the optical microscope. The top and bottom hBN thickness is choosen to be $\sim 10-12 nm$ and $\sim 20-30 nm$, respectively. After exfoliation, the flakes are picked up [2] in appropriate order at $100°C$ from the substrate using the PDMS/PC stamp and a homemade heater stage. Using a micro-manipulator equipped with high accuracy x,y, and z axis movement the flakes are carefully stacked under the microscope. After completing the pick-up sequence, the stacks are dropped on a fresh $Si/SiO_2$ substrate at $180°$ C along with polymer. The polymer is dissolved in solvents (eg. chloroform) and the substrate with the stack is further cleaned with acetone/IPA.

InAs nanowires (NW) are grown on (111)B substrate by the Molecular Beam Epitaxy (MBE) method [3,4]. The NWs have typical diameter ranging from 50 nm to 80 nm. The NWs are first dispersed into ethanol

---

*anindya@iisc.ac.in

solution by sonicating a small piece of the NW grown substrate with the ethanol for few seconds. The NWs are spread by drop-casting 2-5$\mu l$ of the ethanol solution on another clean $Si/SiO_2$ substrate with pre-defined alignment marks. As the ethanol evaporates, the NWs sit on the substrate due to the van der-Waals attraction force. NWs with diameter around 70-80 nm (identified by SEM imaging) are picked up from the substrate by the same dry pick-up technique using a PDMS/PPC stamp at room temperature and subsequently dropped over the clean hBN/Graphene/hBN stack at around $60-70°C$ after careful alignment such that the nanowire is parallel to the graphene edge. The polymer (PPC) is dissolved in solvents and the heterostructures are further cleaned by IPA/Acetone thoroughly.

## SI-1B. Details of contact fabrication

After assembling the heterostructures, the samples are spin coated with bilayer (495A4/950A4) PMMA (e-beam resist) and baked at $180°C$ at the hotplate. The contacts for graphene and NW are made separately. Each time we follow the similar process of spin-coating, baking followed by contact patterning with e-beam lithography. To establish 1D contacts to the encapsulated graphene we follow the well-known reactive-ion etching technique [5] and subsequent thermal evaporation of Cr (5 nm)/Pd (13 nm)/Au (70 nm). The InAs NWs have a thin native oxide layer on their surface which is removed in order to establish ohmic contact. We use the standard technique of chemical etching using 0.3% $(NH_4)_2S$ solution for 30 minutes at $40°C$ [6] and quickly load the samples in the evaporation chamber for thermal evaporation of Ti (5 nm)/Al (100 nm). Before the chemical etching, we perform $O_2$ plasma for few seconds which helps to achieve better contacts to the NWs. The optical and the SEM images of the Coulomb drag devices are shown in Fig. S-1A (a) and (b). The typical channel length of the graphene and the nanowire are respectively $\sim 10\mu m$ and $\sim 1\mu m$. After the fabrication, the samples were cut and mounted on a chip carrier for wire bonding. All the samples were carefully checked at first at room temperature and then at 1.5K for further characterization.

For the Coulomb drag study, the thickness of the top hBN is very important, as it determines the separation between the drive (graphene) and the drag (nanowire) layer. Too thick hBN will reduce the inter-layer Coulomb scattering and may lead to smaller or negligible drag signal which is difficult to measure, whereas very thin layer of hBN will increase the possibility of interlayer leakage current. Significant amount of inter-layer leakage current is unwanted in this kind of scenario as it can mask the actual signal and lead to other effects [7]. For this reason, we limit the top hBN thickness in our devices to be $\sim 10-12nm$ (Fig. S-1A (c)).

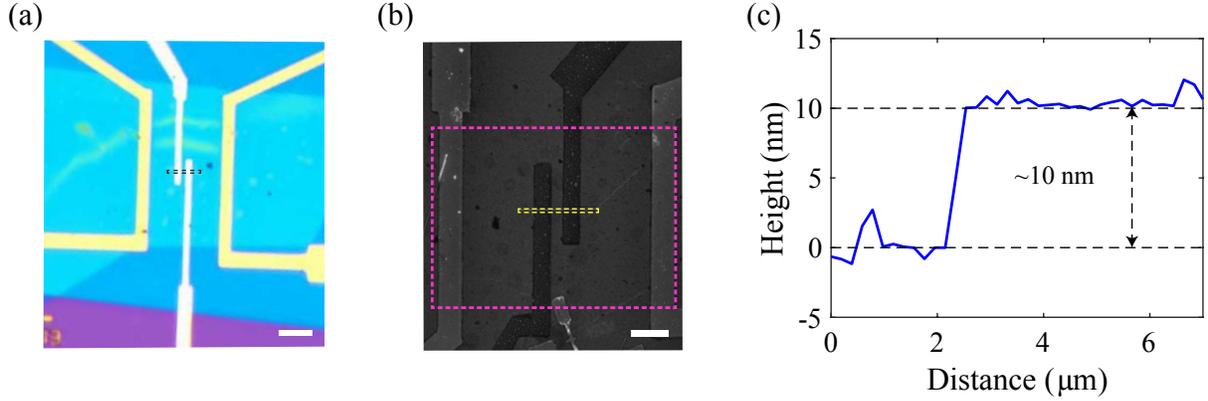

**Fig. S-1A: Device fabrication.** (a) Optical image of a conventional Coulomb drag device. The black dashed line represents the InAs NW. The outer and inner metal contacts are for graphene and NW respectively. The scale bar is 5 $\mu m$. (b) Scanning electron microscope image of a Coulomb drag device. The scale is 2 $\mu m$. The pink dashed line outlines the area of the graphene flake, whereas the yellow dashed line borders the position of the NW which sits on top of the encapsulated graphene stack. (c) AFM height thickness profile of the top hBN of the encapsulated graphene stack showing $\sim 10 nm$.

### SI-1C. Device characterization of Graphene

In this sub-section we will discuss about the extraction of the device parameters such as the density inhomogeneities ($\delta n$), mobilities for the MLG and BLG. We will also discuss how the carrier density of graphene ($n_G$) have been converted from the gate voltage $V_{BG}$. Fig. S-1B and S-1C represent the device characteristics of two MLG-NW and two BLG-NW devices, respectively. Fig. S-1B(a) and (d) show backgate responses of two MLG devices (D1 and D2). Similar plots for BLG devices (D3 and D4) are shown in Fig. S-1C(a) and (d). All four plots have inset images where 2-probe graphene resistance $R$ is plotted against $n_G$ and the plots are fitted with the formula $R = 2R_C + \dfrac{L/W}{e\mu_{FE}\sqrt{n_G^2 + \delta n^2}}$ to extract the field-effect mobility $\mu_{FE}$, where $R_C$, $L$, $W$ are contact resistance, length and width of the graphene channel, respectively. $\mu_{FE}$ is found to be $\sim$ 100,000 and 60,000 $cm^2/VS$ for D1 and D2, whereas 65,000 and 53,000 $cm^2/VS$ for D3, D4, respectively. To find the density inhomogeneities in graphene, the graphene conductance $G$ is plotted against $n_G$ in the log-log fashion as shown in Fig. S-1B (b), (e), and S-1C (b),(e). We obtain $\delta n$ to be around $2 - 3 \times 10^{10}/cm^2$ which is standard for encapsulated graphene devices [8,9].

The graphene density ($n_G$) is tuned by the back gate voltage $V_{BG}$. We apply the backgate voltage $V_{BG}$ in the p++ doped $Si$ substrate across the 300 nm thick thermal oxide. Applying $V_{BG}$ would dope

3

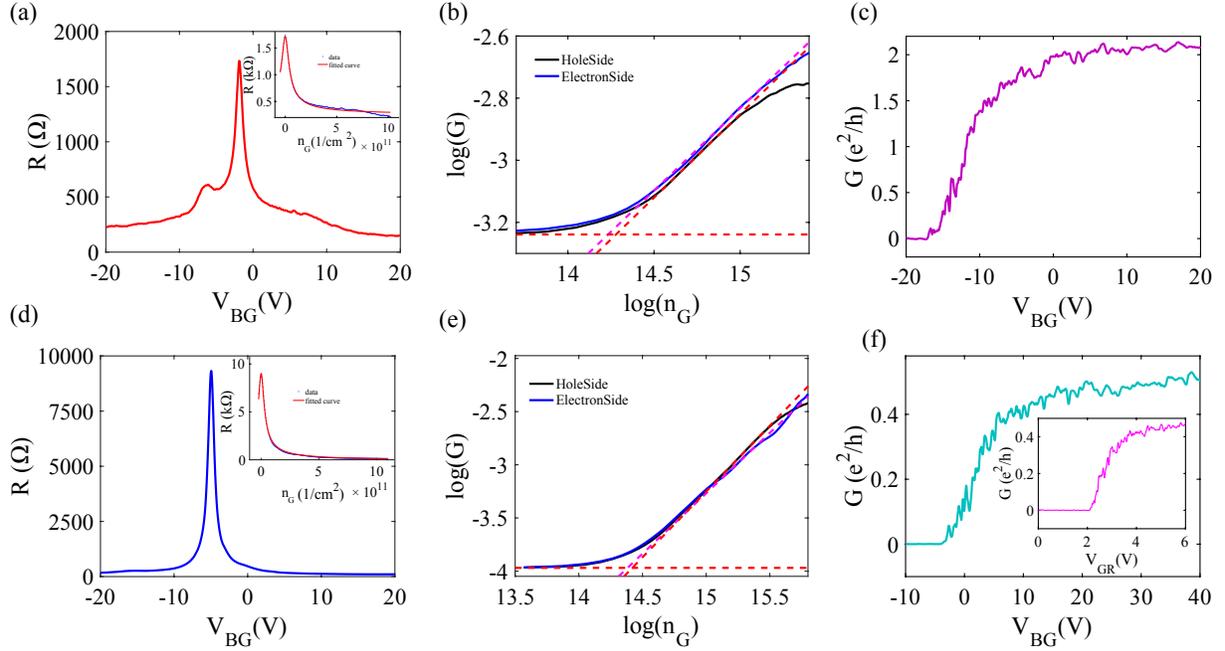

**Fig. S-1B: Monolayer Graphene-NW (MLG-NW) device characterization.** (a) and (d) 2-probe resistance of MLG-NW devices (D1 and D2) is plotted against backgate voltage $V_{BG}$ at T=1.5K. Inset shows the gate response of the respective devices fitted with the mobility formula with mobility $\mu_{FE}$ and contact resistance $R_C$ as fitting parameters. The extracted mobilities are found to be $\sim 100,000\ cm^2/VS$ and $\sim 60,000\ cm^2/VS$, respectively for D1 and D2. In (b) and (e) the intrinsic inhomogeneity ($\delta n$) have been extracted by plotting $\log G$ vs. $\log n_G$ for the D1 and D2 devices, respectively for electron as well as hole side. The extracted $\delta n$ values are $1.71(1.9) \times 10^{10}/cm^2$ and $2.41(2.63) \times 10^{10}/cm^2$ for electron (hole) side in D1 and D2, respectively. (c) and (f) are backgate responses of 2-probe conductance of the InAs NW at 1.5K while graphene contacts are in floating condition for D1 and D2 devices, respectively. Inset panel in (f) is the NW conductance plotted with $V_{GR}$.

the graphene. The carrier density per unit area accumulated on the graphene can be expressed as: $n_G = \frac{1}{e}C_G(V_{BG} - V_{DP})$, where $V_{DP}$ is the backgate voltage where the graphene becomes charge neutral. The quantity $C_G$ is the effective capacitance per unit area between the graphene and the doped $Si$, given by $C_G = \frac{\epsilon_0 \epsilon_r}{d}$, where $\epsilon_0 = 8.854 \times 10^{-12} Fm^{-1}$ is the permittivity of free space, $\epsilon_r = 3.9$ is the relative permittvity of $SiO_2$ and $d = 300nm$ is the thickness of the oxide layer. By putting all the parameters, we obtain $C_G = 115 aF\mu m^{-2}$.

Apart from using optical microscope, we have employed Quantum hall (QH) plateaus to identify monolayer and bilayer graphene in the heterostructures. Fig. S-1D (a) and (b) shows the the characteristic

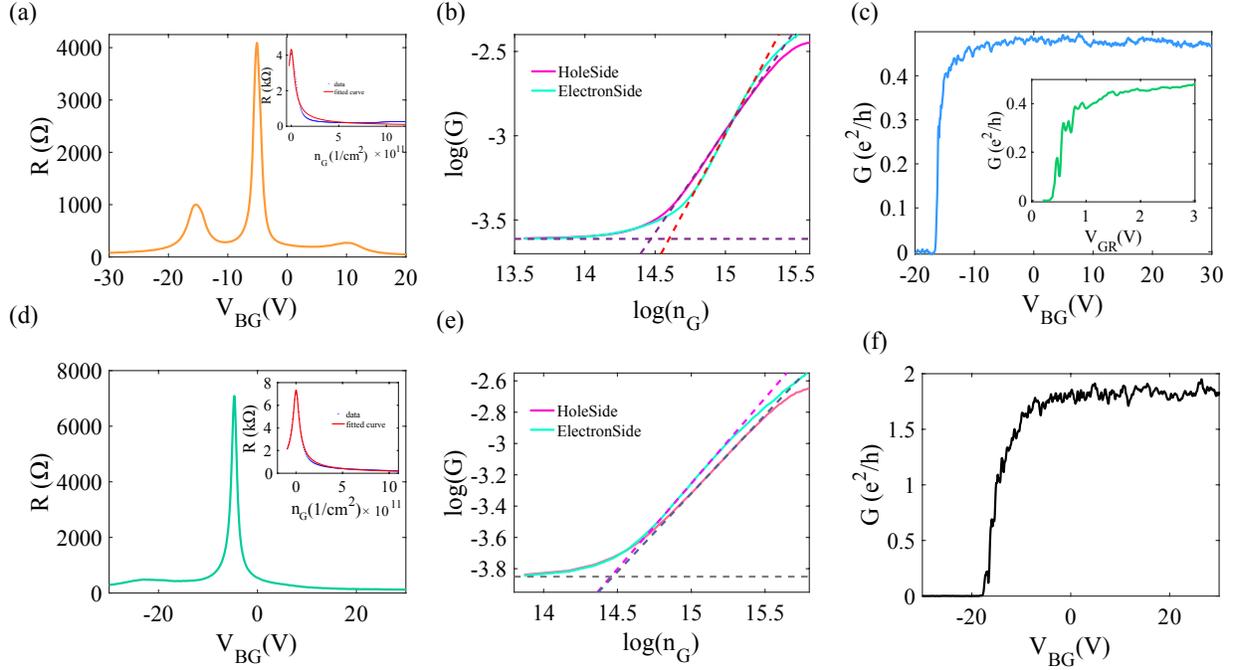

**Fig. S-1C : Bilayer graphene-NW (BLG-NW) device characterisation.** (a) and (d) 2-probe resistance of BLG devices D3-BLG and D4-BLG plotted against backgate voltage $V_{BG}$ at T=1.5K. Inset shows the gate response fitted with the mobility formula with mobility $\mu_{FE}$ and contact resistance $R_C$ being the fitting parameters. D3 and D4 have mobilities $\sim 65,000\ cm^2/VS$ and $\sim 53,000\ cm^2/VS$, respectively. (b) and (e) the intrinsic inhomogeneities ($\delta n$) of the BLG devices are calculated by plotting $\log G$ vs. $\log n_G$. $\delta n \sim$ 2.95 (3.97) and 2.85 (2.95) $\times 10^{10}/cm^2$ for electron (hole) side for D3 and D4, respectively. (c) and (f) are backgate response of 2-probe conductance of the InAs NWs for D3 and D4 devices respectively at 1.5K while graphene contacts are in floating condition. Inset panel is the NW conductance plotted with $V_{GR}$.

QH plateaus for MLG and BLG.

## SI-1D. Device characterization of InAs Nanowires

InAs is a n-type semiconductor with direct bandgap of 0.35 eV. Due to fermi level pinning within the conduction band, the nanowires are always n-type. The 2-probe conductance ($G$) versus $V_{BG}$ (while the graphene is at floating condition) corresponding to devices D1, D2, D3 and D4 are shown in Fig. S-1B(c),(f) and S-1C(c),(f), respectively. All the nanowires start conduction at a threshold voltage ($V_{TH}$). Below $V_{TH}$ the conductance remains negligible which gradually increases and become constant (determined by

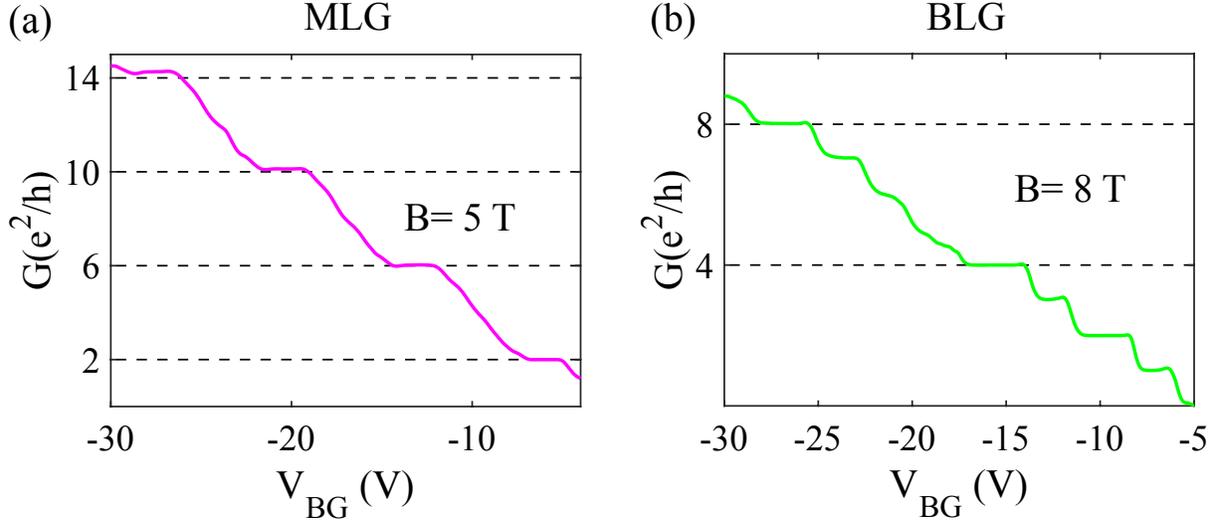

**Fig. S-1D: Quantum Hall plateaus in MLG and BLG**. (a) and (b) Quantum Hall response of D1-MLG and D3-BLG devices at 1.5K and captured at 5T and 8T magnetic field, respectively. (a) shows characteristic conductance plateaus of the MLG which appears at conductance of 2, 6, 10 $G_0$, whereas (b) shows conductance plateaus of BLG with 4, 8 $G_0$ conductance plateaus.

the contact resistance) at higher gate voltages. Since for most of the devices we obtain $V_{TH} < 0$, the nanowires show significant conduction even at $V_{BG} = 0$ due to electron doping. During the Coulomb drag measurement one of the graphene contacts is kept grounded. To change the carrier density ($n_N$) of the nanowire in this situation, we apply a voltage $V_{GR}$ between the graphene and the nanowire through the top hBN, as discussed in the measurement technique section. Fig. S-1B(f) and S-1C(c) inset show the nanowire conductance as a function of $V_{GR}$. For calculating $n_N$ from $V_{GR}$, we use $n_N = \frac{1}{e}C_N(V_{GR} - V_{TH})$, where $C_N$ is the capacitance per unit length. In order to evaluate $C_N$, we use the cylinder on a infinite plate capacitance model [10,11], where $C_N = \frac{2\pi\epsilon_0\epsilon_r}{\cosh^{-1}(t/r)}$, where $t$ is the distance between the center of the nanowire to the graphene, and $r$ is the radius of the NW. For our devices, the top hBN thickness $\sim 10nm$ and $r \sim 40nm$ and thus $t \sim 50nm$. Putting all the parameters, we obtain $C_N = 320 aF\mu m^{-1}$.

**Mobility calculation of the Nanowire:** We calculate the field-effect mobility of the InAs nanowires using the analytical expression [12]:

$$\mu = \frac{L}{C_N}\frac{dG}{d(V_{GR} - V_{TH})} \tag{1}$$

Where, L is the channel length, $C_N$, $V_{GR}$ and $V_{TH}$ are the capacitance per length, gate voltage and the threshold voltage respectively as mentioned earlier. The term $\frac{dG}{d(V_{GR}-V_{TH})}$ is calculated from the slope of the gate response of the nanowire 2-probe conductance. For most of the nanowires used in the Coulomb

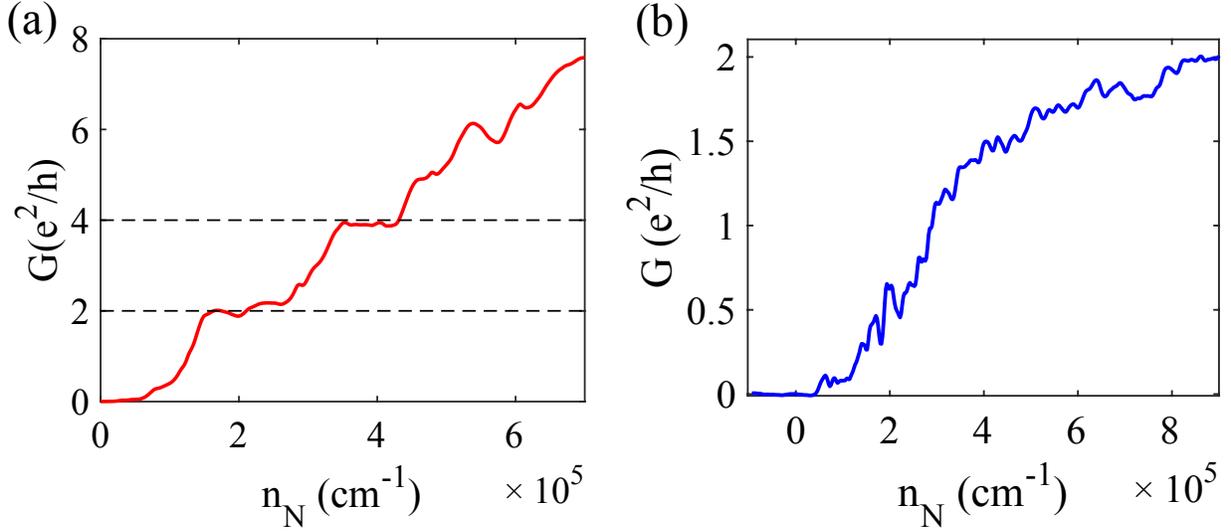

**Fig. S-1E : 1D nature of the InAs nanowires.** (a)Conductance of InAs nanowires as a function of carrier concentration for 100 nm and (b) 1 $\mu m$ channel lengths.

drag experiments, we obtain mobilities to be $\sim 2000 - 3000 \ cm^2/V.s$.

### SI-1E. The 1D nature of the InAs nanowires

In order to establish that the InAs nanowires used in our studies are indeed 1D systems, we have measured the conductance as a function of gate voltage for devices having two different channel lengths i.e. $\sim 100$ nm and $\sim 1 \ \mu m$, as shown in Fig SI-1E. The 100 nm channel device ( Fig. SI-1E a) clearly shows the signature of different 1D sub-bands by exhibiting conductance plateaus at $2\frac{e^2}{h}$ and $4\frac{e^2}{h}$. It can be seen that within the density range of $\sim 7 \times 10^5 cm^{-1}$ (range accessed in our experiments), maximum of five sub-bands are populated. In comparison, the 1 $\mu m$ channel length shows monotonic increase of the conductance (Fig. SI-1E b) with gate voltage and saturates around $2\frac{e^2}{h}$. This suggests that the transport in $\sim 1 \ \mu m$ nanowire channel (used in our drag experiments) is not in the ballistic regime but rather in diffusive regime with mobility around $\sim 2000 cm^2/V.s$ as mentioned in the previous section. The above arguments also justifies that the InAs nanowires used in our experiments are not heavily doped, rather only few sub bands are populated. The conductance value for 1 $\mu m$ channel length is limited due to its diffusive nature.

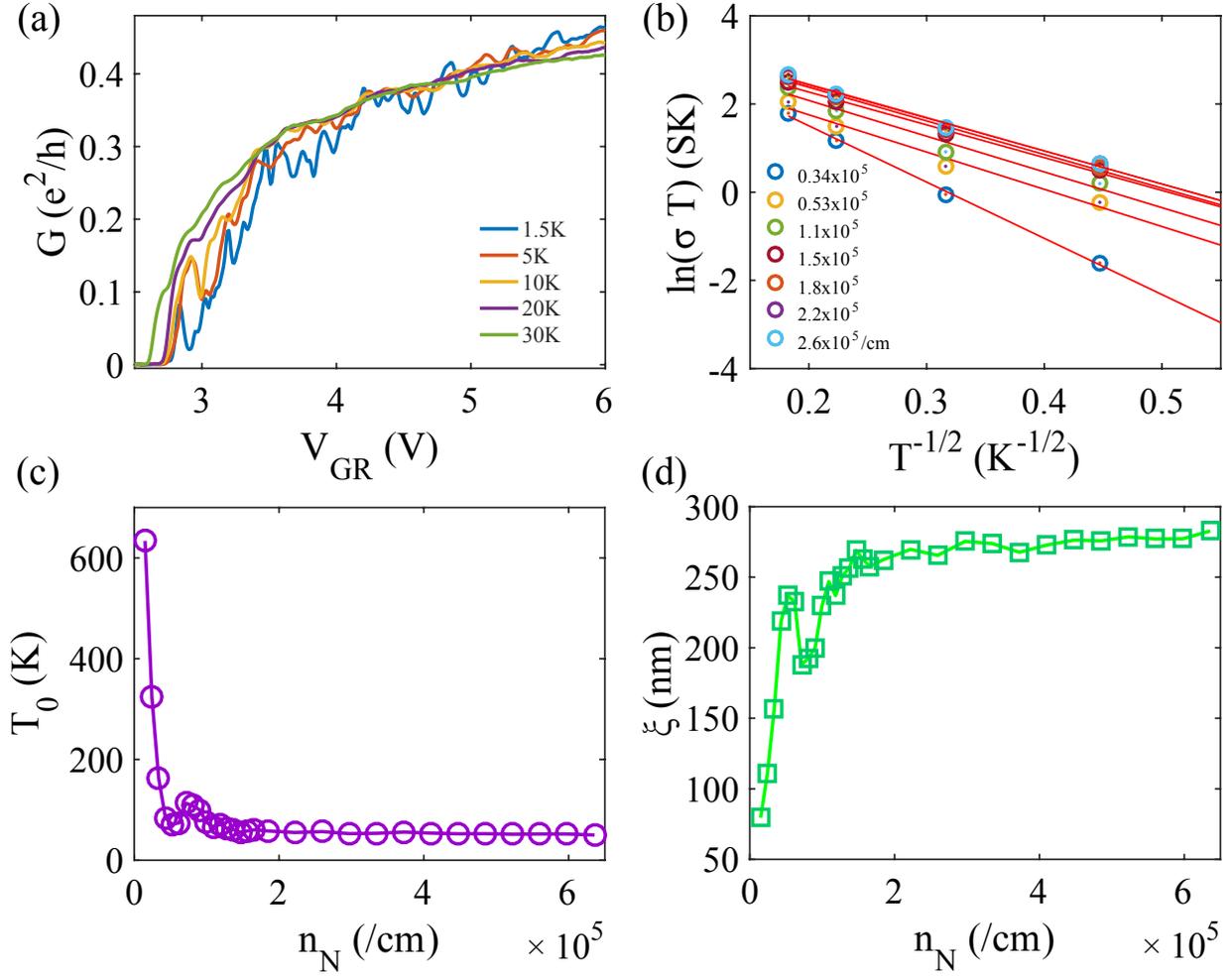

**Fig. S-1F : Localization length ($\xi$) extraction.** (a) 2-probe conductance of the NW (D2-MLG) plotted with $V_{GR}$ at different temperatures from 1.5K to 30K. The conductance oscillations present at low temperature disappear at the high temperature regime. (b) Localization length ($\xi$) is extracted by fitting $\sigma$ at different T according to Mott formula of variable range hopping. $log(\sigma T)$ with $T^{-1/2}$ for different $V_{GR}$ gives $T_0$. The open circles and the red solid lines are respectively data and fitted lines for multiple $n_N$. (c) Extracted $T_0$ plot with 1D NW densities $n_N$. (d) Localization length ($\xi$) plotted with 1D NW density $n_N$.

8

## SI-1F. Localization length ($\xi$) in InAs Nanowires

The localisation length ($\xi$) in the InAs NWs has been calculated using the Mott formula from the temperature dependence of the conductivity $\sigma$ at different nanowire densities. By calculating $\xi$ we can estimate the length scale of the puddles present in the NW. The Fig. S-1F (a) shows the gate response of InAs NW (belongs to D2 device) at different temperatures. Here we have modeled the NW transport by variable range hopping (VRH), where the conductivity $\sigma$ of the d dimensional system at temperature T is expressed as [13]:

$$\sigma = \sigma_0(T)\ exp[-(T_0/T)^{\frac{1}{d+1}}] \tag{2}$$

where $T_0$ and d are correlation energy scale and dimensionality of the system respectively, and $\sigma_0 = AT^m$, $m \approx 0.8 - 1$ [13]. We have extracted the $T_0$ by plotting $ln(\sigma T)$ vs. $T^{-1/2}$ which corresponds to d=1 of equation (2), and measuring the intercept while linearly fitting the data (red lines in Fig. S-1F (b). In Fig. S-1F (c), the $T_0$ is plotted with the 1D NW densities $n_N$ which shows the energy dependence of $T_0$. To extract $\xi$ from $T_0$, we use $\xi^2 = \dfrac{13.8}{k_B T_0 D(E)}$, where $D(E) \sim 4 \times 10^{12} eV^{-1} cm^{-2}$ [14] is the typical surface density of charged trap at oxide substrate. In the VRH model, we consider the electron transport occurs via band of localized disordered states [13]. Figure S-1F(d) shows $\xi$ plotted against $n_N$ where $\xi \sim 100 - 200 nm$.

## SI-2: Measurement technique

In Coulomb drag, a constant current ($I_D$) is passed through the drive layer and as a result of inter-layer momentum and energy transfer, an open circuit voltage ($V_D$) is generated in the drag layer without any exchange of particle. Now, Coulomb drag for the MLG/BLG-NW devices can be measured in two configurations: (a) driving current in graphene and measuring voltage in NW, (b) driving current in NW and measuring voltage in graphene.

## SI-2A. DC vs. AC measurement

In Coulomb drag measurements, both DC and AC techniques have been utilized to measure the drag voltage ($V_D$). The measurement schematic for the DC and AC technique are presented in Fig. S-2A(a) and S-2A(b),

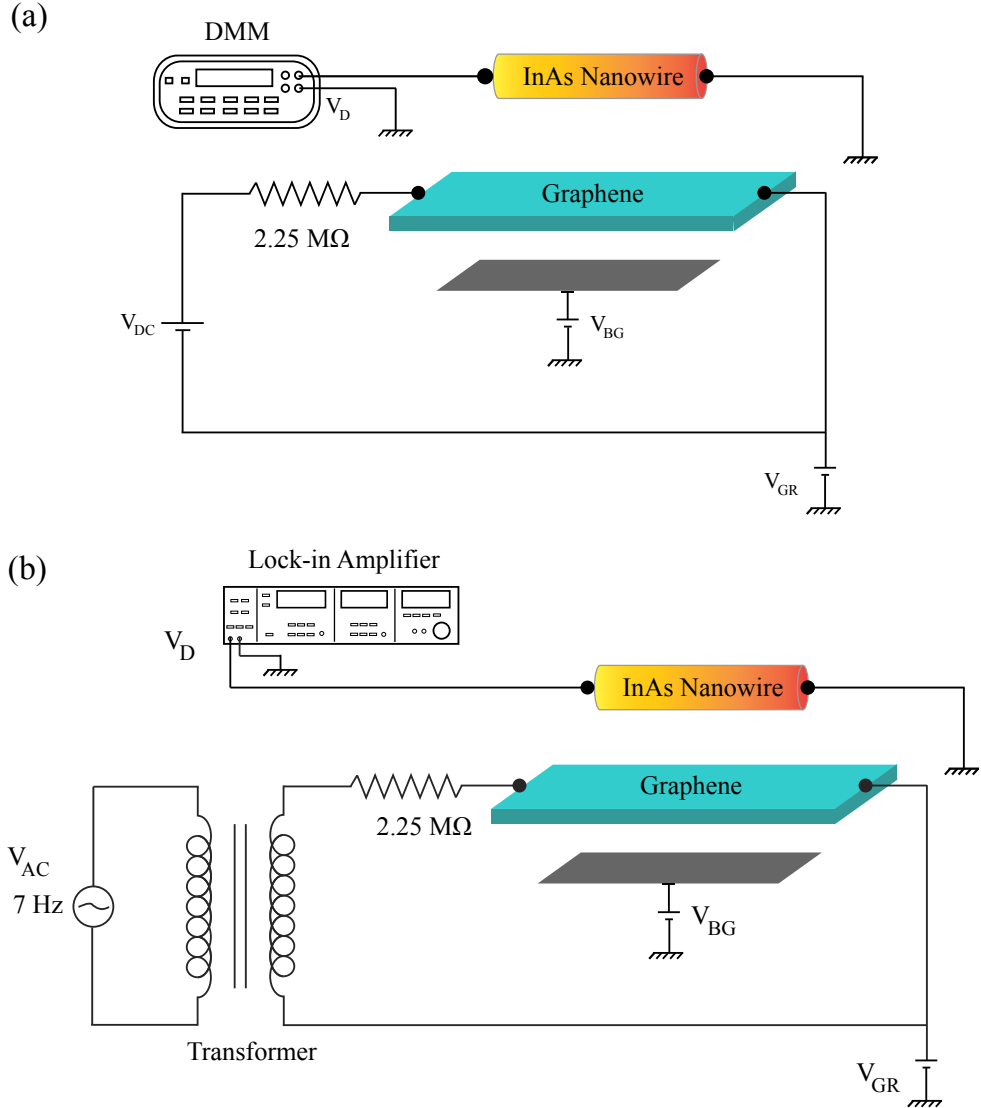

**Fig. S-2A: DC and AC measurement schematic.** Coulomb drag measurement schematic for (a) DC and (b) AC technique. In DC technique, (a) a DC voltage is applied across the drive layer through a large ($\sim M\Omega$) resistance and the open circuit voltage appearing in drag layer is measured by a digital multimeter (DMM) having the input impedance of 100M$\Omega$. Backagte voltage $V_{BG}$ is applied in Si through $SiO_2$ for changing graphene density $n_G$, while NW is maintained at a constant conductance. An additional voltage $V_{GR}$ is applied in graphene which shifts the fermi energy of the NW with respect to graphene. For the AC technique, (b) a small frequency ($\sim$ 7 Hz) AC signal from the Lock-in amplifier is applied across the drive layer through a high resistive element and then the open circuit AC voltage is measured through Lock-in. The gate voltage $V_{GR}$ is connected with the circuit using a isolation transformer as shown.

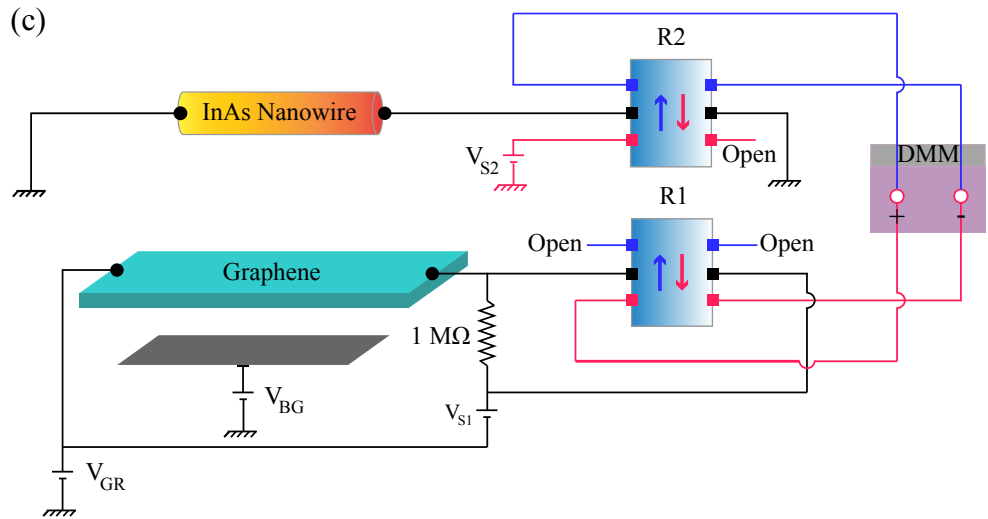

**Fig. S-2A: Double Relay circuit.** (c) We have performed DC measurement using this circuit to confirm the Onsager reciprocity relation. Each relay R1 and R2 are connected to the two layers separately as shown. The circuit is connected such that one can independently interchange between the graphene and NW layers as drive layers without changing the $V_{GR}$ (NW density). The blue (red) color lines indicate where graphene (NW) is the drive layer and the NW (graphene) is the drag layer. $V_{S1}$ and $V_{S2}$ is the DC source voltage connected to the graphene and NW, respectively.

respectively. In the DC technique, a DC voltage from a voltage source (Keithley 2400 or Yokogawa GS200) is applied across the drive layer (Graphene) through a high resistance ($\sim 1 - 3M\Omega$) path, and the open circuit output voltage is measured across the drag layer (nanowire) by a Digital multimeter (Agilent 34401A 6 1/2 DMM). For the AC technique, a low frequency ($\sim 7Hz$) AC signal from the output of SR 830 lock-in amplifier is applied across the drive layer whereas the output drag signal is fed into the lock-in amplifier. The input voltage is changed such that $I_D$ varies from + $10\mu A$ to - $10\mu A$. In order to change the nanowire density, we apply $V_{GR}$ through graphene. For the AC circuit (Fig. S-2A (b)), an isolation transformer circuit is used for applying the DC voltage $V_{GR}$. Since drag voltage is very sensitive to the carrier density of the NW, we have kept $V_{GR}$ fixed at a certain value while measuring the $V_D$ verses $V_{BG}$. We have applied a circuit (Fig. S-2A(c)) using two relays such that the drag can be measured in both the configurations. By switching the relays in appropriate manner, we are able to measure the $V_D$ for multiple NW densities $n_N$ in both the configurations. It has helped us to investigate the validity of the Onsager principle appropriately.

## SI-2B. Flipping and non-flipping part extraction

Although both DC and AC techniques (section **SI-2A**) yield same drag features (Fig. SI-2B (d)), the raw signals in DC measurement are interpreted in a different way. We have observed that, the raw DC signal contain predominantly a drag signal superposed with a small non-flipping signal. The drag signal (flipping part) flips sign when the current direction is flipped whereas the sign of the non-drag signal (non-flipping part) remains unchanged. In order to extract the actual drag signal from the raw data, we use the following protocol: $V_D \to -V_D$, as $I_D \to -I_D$ for drag signals, but the non-flipping part which originates from heating effect ($\propto I^2 R$) doesn't changes its sign. So, we can write in equations that:

$$\begin{aligned} V^+{}_{raw} &= V_{FP} + V_{NFP} \\ V^-{}_{raw} &= -V_{FP} + V_{NFP} \end{aligned} \quad (3)$$

where $V^+{}_{raw}$ and $V^-{}_{raw}$ are raw drag signals when $I_D$ is positive and negative, respectively. $V_{FP}$ and $V_{NFP}$ are contributing flipping and non-flipping part of the drag signal, respectively. Combining the two equations, we get:

$$V_{FP} = \frac{1}{2}(V^+{}_{raw} - V^-{}_{raw}) \quad (4)$$

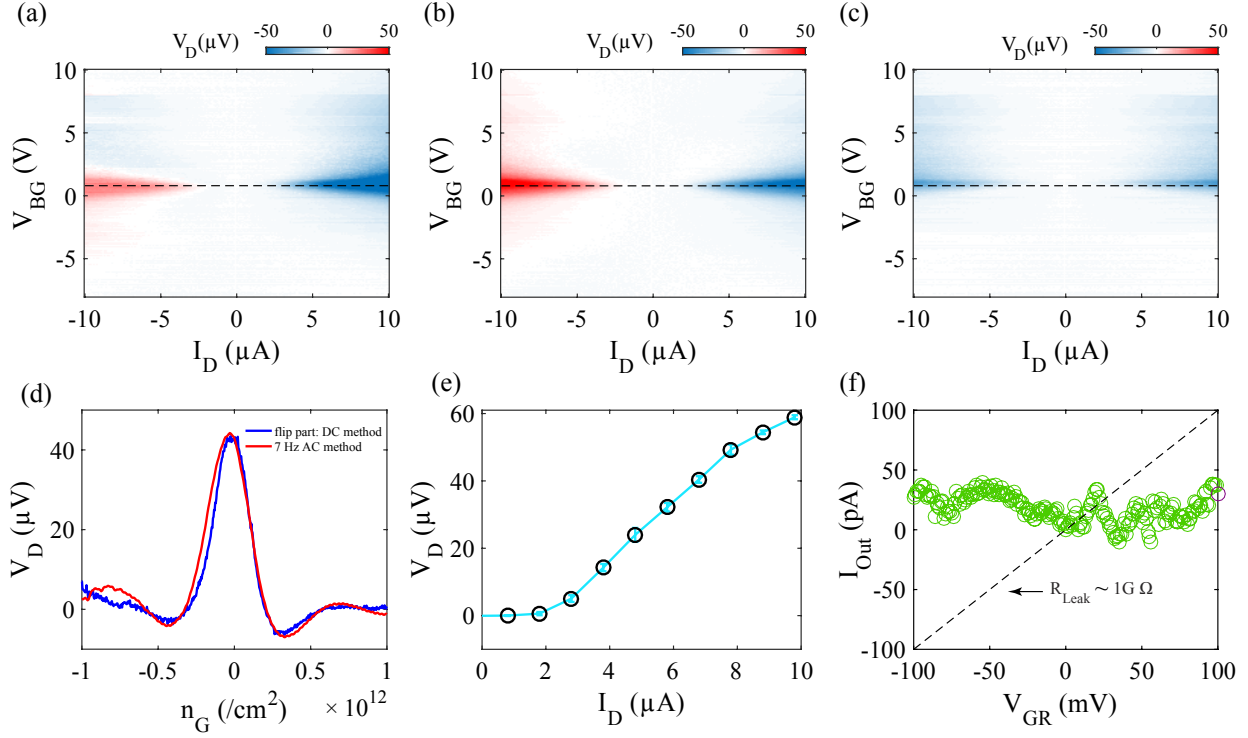

**Fig. S-2B: Flipping and non-flipping part extraction.** (a) 2D colormap (raw data) of open circuit drag voltage plotted against drive current $I_D$ along the x-axis and backgate voltage $V_{BG}$ along the y-axis for a MLG-NW device at 1.5K. (b) and (c) are similar 2D colormaps of the extracted flipping and non-flipping part of the drag signal as mentioned in the text. In (b), the drag is negative for positive $I_D$ and flips sign as the current direction is flipped unlike (c). Color-bar ranges are same for plots of (a)-(c) and the black dashed line indicates the Dirac point of the system. (d) Comparison between drag voltage acquired in DC and low frequency AC technique at 1.5K for a MLG-NW device. (e) $V_D$ plotted against the $I_D$ extracted from the 2D plot. $V_D$ varies quite linearly with $I_D$. (f) Inter-layer leakage current ($I_{Out}$) is plotted against the inter-layer voltage ($V_{GR}$) applied across the graphene and the NW. The black dashed line corresponds to the leakage resistance $R_{Leak} \sim 1 G\Omega$. Our data implies $R_{Leak}$ in our device is much larger than $1 G\Omega$.

13

All the data presented in the main text are flipping part and extracted using equation (4). Fig S-2B(a) is the raw data collected by DC technique for a MLG-NW device, and S-2B(b) and S-2B(c) are the flipping and non-flipping part extracted by the above mentioned process. We can observe that, the non-flipping part is much smaller in magnitude compared to the drag signal. To cross-check the validation of this extraction process, we compare the extracted flipping DC data with the raw AC drag signal and we have found that they match remarkably well with each other as shown in Fig S-2B(d). Figure S-2B(e) shows the drag signal as a function of drive current, which varies linearly beyond $2\mu A$ which also ensures that all the data has been recorded while the system is in the linear regime. Figure S-2B(f) shows the leakage current through the top hBN as a function of $V_{GR}$. One can clearly see that the leakage resistance is much larger than $1G\Omega$.

As seen from Fig. S-2B(b) and Fig. 1(b), 2(a) of the main manuscript, the drag signals (flipping part) in our samples have magnitude comparatively smaller ($\sim 1\Omega$) than the well-studied 2D-2D systems reported so far. In a dimensional mismatched system, smaller drag is expected due to the limited phase space involved in scattering as compared to the 2D-2D systems. In hybrid systems like ours, only a fraction of the drive current can interact with the carriers in the other layer to produce the Coulomb drag. In a simplified picture, the drag resistance will be proportional to the ratio of width of the drag layer to the width of the drive layer ($\frac{W_{Drag}}{W_{Drive}}$). In 2D-2D system this ratio is unity, whereas in 2D-1D hybrid the ratio is two orders smaller. Therefore, the observed drag resistance in our case ($\sim 1\Omega$) roughly scales to $\sim 100-200$ $\Omega$ for the 2D-2D devices, close to the observed values [15].

Although measuring small drag signal ($\sim 1\Omega$) was challenging but we could measure it accurately as the drag voltage was few tens of micro volts, which was much higher than the resolution of 100 nV.

### SI-3A. Tuning $n_G$ and $n_N$

In the Coulomb drag measurements, the drag resistance $R_D = \frac{V_D}{I_D}$ has been captured as a function of both $n_G$ and $n_N$. Applying $V_{BG}$ in doped Si through the $SiO_2$ tunes the $n_G$. In order to change $n_N$, we apply $V_{GR}$ to the graphene layer as shown Fig. S-2A. The following two equations [16] demonstrate how the carrier densities change with the gate voltages:

$$C_G (V_{BG} - V_{GR}) = n_G e \quad (5)$$
$$C_N V_{GR} = n_N e \quad (6)$$

14

where $C_G$ and $C_N$ are the capacitance per unit area and capacitance per unit length, respectively between graphene sheet and the p-doped Si with $SiO_2$ as the dielectric medium, and between the cylindrical shaped NW and the graphene sheet where the top hBN of the heterostructure acts like a dielectric medium. The quantities $n_G$ and $n_N$ are 2D and 1D carrier densities of the graphene and the nanowire, respectively. In the equations above, we have not taken into account the effect of quantum capacitance of the layers, as that doesn't affect the qualitative outcome of our results.

### SI-3B. Shifting of Dirac point with application of $V_{GR}$

The 2-probe resistance versus backgate voltage $V_{BG}$ response of the MLG of the D2 device at T=1.5K for different $V_{GR}$ values is shown in Fig. S-3A. The shift of the Dirac point towards more electron-side is governed by equations (4) and (5). We have taken into account this effect for cases where $V_{GR} \neq 0$ while presenting $n_G$.

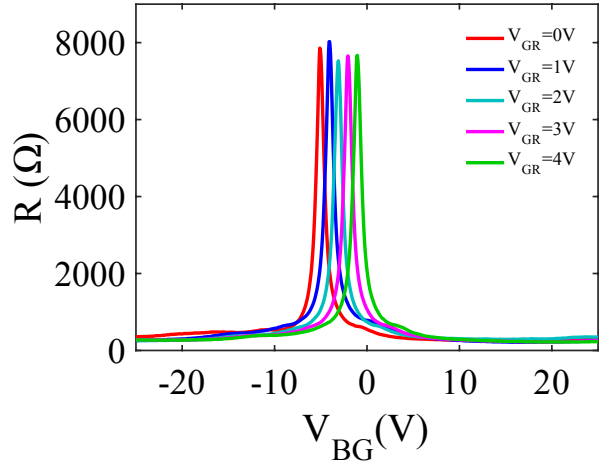

**Fig. S-3A: Dirac point shift.** Backgate response of graphene 2-probe resistance of D2-MLG at 1.5K for different graphene gate voltage $V_{GR}$. With increasing $V_{GR}$, the Dirac point of graphene shifts towards positive backgate voltage as the graphene becomes more hole type doped for $V_{GR} > 0$.

## SI-3C. Drag peak dependence on Nanowire carrier density for MLG-NW device

In this section we present the data showing variation of $R_D$ with the $n_N$. The conversion of $n_N$ from $V_{GR}$ has been mentioned before in section SI-1D where it is shown that the threshold voltage value $V_{TH}$ of the device can directly influence $n_N$. Evidently, any error in obtaining the $V_{TH}$ can lead to an error estimating $n_N$ from $V_{GR}$ and further the density dependence of the drag signal. As shown in Fig. S-3B (a), the gate responses of the nanowires in our devices are quite reproducible and we could determine the threshold voltage quite accurately from the $logG$ versus $V_{BG}$ plot. However, from Fig. S-3B (a) we see that there is small difference in the threshold voltages between two successive gate voltage sweeps. We can quantify the error in estimating $n_N$ as $\delta n = C_N \delta V_{BG} \sim 0.38 \times 10^5 cm^{-1}$, where $\delta V_{BG}$ is the spread between the threshold voltages of different sweeps. Fig. S-3B(b) shows the drag resistance ($R_D$) response with the graphene density ($n_G$) at different values of $n_N$ for a MLG-NW device. The drag resistance peak decreases as the $n_N$ increases. The $R_D$ peak magnitude appearing at $n_G = 0$ is plotted against $n_N$ in Fig. S-3B (c)(also shown in Fig. 1(g) of the main text). We have also included the error in estimating $n_N$ as horizontal error bars shown in Fig S-3B (c). The red solid line shows the agreement of our data with $R_D \sim n_N^{-4}$.

From fig S-3B(b), we notice that for certain nanowire densities, the drag resistance becomes negative in the intermediate $n_G$ values (for $n_N = 1.79 \times 10^5 cm^{-1}$). Although we don't have a clear understanding of this negative $R_D$ at finite $n_G$, We believe that the answer may lie in the dimensionality mismatched 2D-1D system. However, for the temperature and magnetic field data presented in the manuscript shown in Fig 1(b)-1(f), has been measured for $n_N \sim 4 \times 10^5 cm^{-1}$ where $R_D$ is positive for all values of $n_G$.

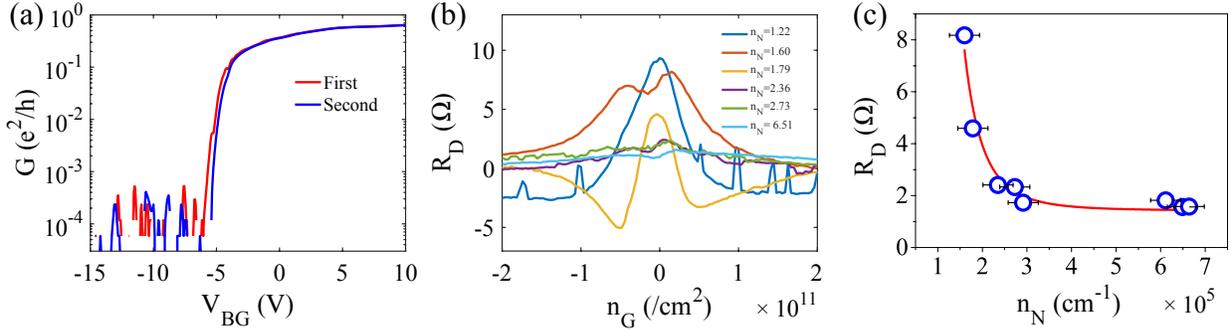

**Fig. S-3B: $R_D$ vs. $n_n$ plot with errorbars.** (a) Nanowire conductance as a function of gate voltages for two successive gate voltage sweeps. (b) $R_D$ plotted against the $n_G$ at T= 1.5K for different $n_N$ in unit of $10^5 cm^{-1}$ controlled by the $V_{GR}$. (c) The $R_D$ magnitude at $n_G = 0$ from (b) is plotted with corresponding $n_N$ with horizontal black errorbar extracted from error in estimating the threshold voltage.

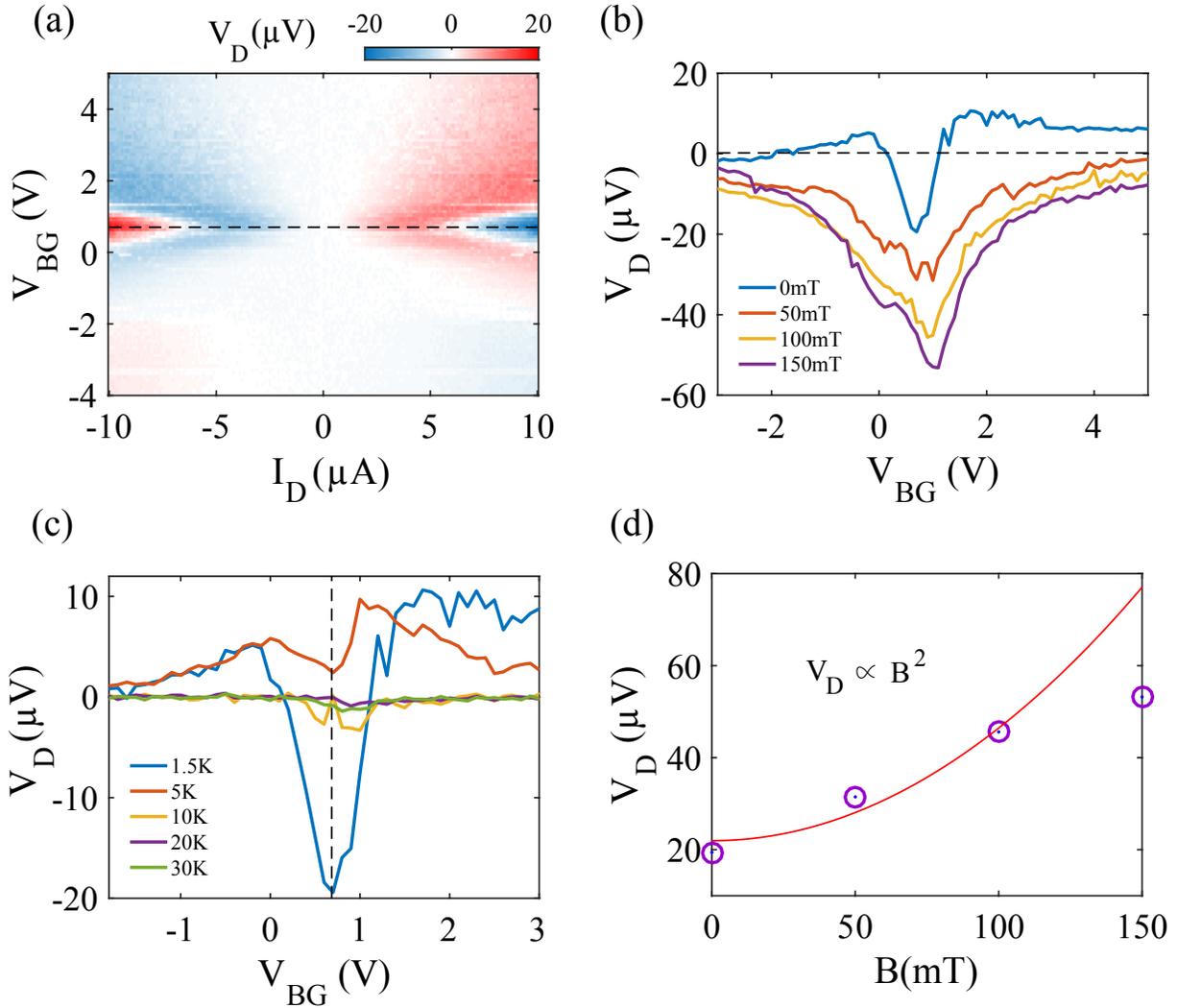

**Fig. S-3C: MLG-NW sample with a dip at the Dirac point.** (a) 2D colormap of $V_D$ (flipping part) plotted against $I_D$ along the x-axis and $V_{BG}$ along the y-axis at T=1.5K. The black dashed line indicates the position of the Dirac point of the sample. At the Dirac point, $V_D$ shows a dip instead of a peak, unlike other MLG-NW samples. (b) Backgate response of $V_D$ at different perpendicular magnetic fields at T=1.5K. $V_D$ rapidly increases in magnitude with magnetic field. (c) Backgate response of $V_D$ at different temperatures. At T=1.5K, $V_D$ shows a central dip at the Dirac point which diminishes quickly at higher temperature. (d) Dip magnitude of $V_D$ plotted with the applied magnetic field. The red line shows $B^2$ fitting of the existing data at low magnetic field.

## SI-3D. MLG-NW device with a dip at the Dirac point

Although most of the MLG-NW devices shows a peak near the Dirac point, for some devices we have observed a dip in drag signal near the Dirac point as shown in Fig. S-2C. The dip has the same characteristics

17

as the peak appearing in other MLG-NW devices. As shown in Fig. S-2C (b) and (c) the $V_D$ increases in magnitude in presence of perpendicular magnetic field and diminishes very quickly with increasing temperature. These data are similar to devices D1 and D2, but with a dip instead of a peak at the Dirac point. Fig. S-2C (d) shows that the dip magnitude plotted with B fits with $B^2$ at smaller values off $B$.

The possible reason that some of the monolayer graphene devices shows a dip instead of a peak, can be related to the type of inter-layer correlation present between the charge puddles. It is known from literature [17–19] that, positive (negative) inter-layer correlation i.e. $\delta\mu_1\delta\mu_2 > 0\ (< 0)$ leads to positive (negative) drag signal due to Energy transfer mechanism. Positive correlation is expected when disorder potential is dominated by charged impurities, whereas puddles due to layer strain often bear negative correlation [17–19].

## SI 4A. Drag signal at different magnetic fields for the BLG-NW device

The raw data corresponding to Fig. 2(d) of the main text showing the dip values of $R_D$ for discrete magnetic values which belongs to the D3 device is presented in this section. As shown in Fig. S-4A, $R_D$ dip magnitude in the electron-side remains almost constant at non-zero magnetic field unlike the MLG-NW devices.

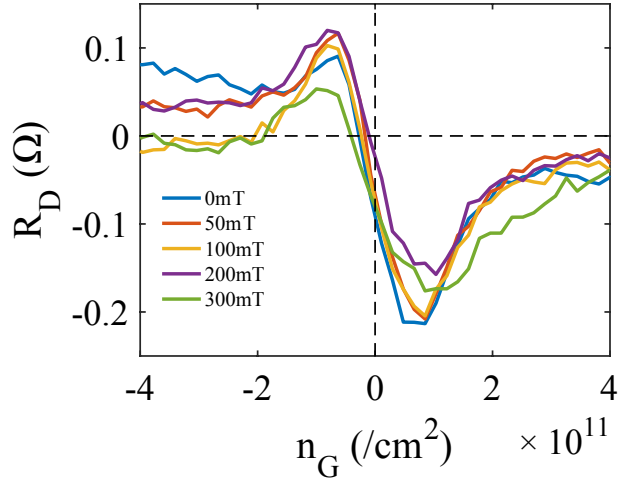

**Fig. S-4A: $R_D$ of BLG-NW device with magnetic field.** $R_D$ vs. $n_G$ is plotted for different magnetic fields for the D3 device at T=1.5K. The $R_D$ doesn't change significantly in presence of the magnetic fields as compared to the MLG-NW devices. The black dashed lines indicate the zero drag magnitude and the $n_G = 0$.

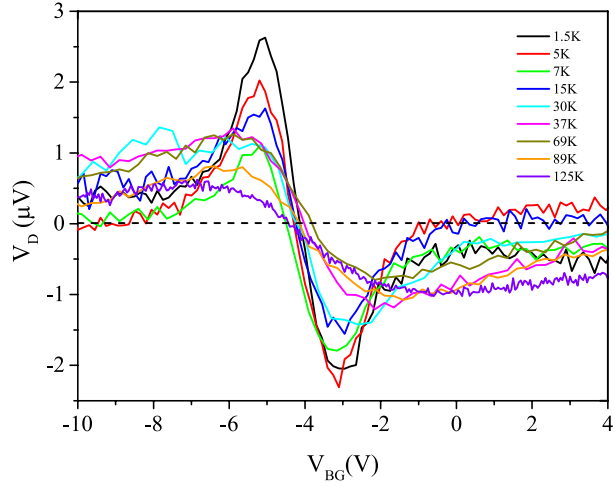

**Fig. S-4B: $R_D$ peak position shift with T.** Backgate response of drag signal for different temperatures plotted together for D3-BLG device at $V_{GR} = 0.9V$. For all temperatures, drag signal flips across the Dirac point. However, at higher temperature the drag signal magnitude diminishes and the peak/dip position of drag shifts towards higher density which affects the allover temperature variation of the drag signal. The horizontal black dashed line is the zero drag magnitude.

### SI-4B: Gate response of Drag signal at different temperatures for the BLG-NW device

In this section, we discuss the anomalous temperature dependence observed for our BLG-NW devices. For the BLG-NW devices, drag follows all the momentum drag properties except that, $R_D$ doesn't increase as $T^2$ with the temperature. Instead the drag peak/dip magnitude reduces slowly with increasing temperature. We also observe that the $n_G$ value at which the peak/dip appears ($n_G^*$), shifts towards the higher value, i.e. $n_G^*$ increases as the temperature rises (shown in Fig. S-4B). The variation of $n_G^*$ and drag magnitude at the peak with temperature have been demonstrated in Fig. 2(e) and (f) of the main text. We believe that, since in Momentum drag the peak/dip position in density is determined by the temperature induced broadening as well as intrinsic inhomogeneities ($\delta n$) of the system, different temperature regime has a role to play. At lower temperature regime when $k_B T < \mu_{\delta n}$ (Here $\mu_{\delta n}$ is the equivalent chemical potential due to intrinsic inhomogeneity $\delta n$), the peak/dip position is determined by the $\mu_{\delta n}$ whereas at higher temperature regime, the temperature induced Fermi energy broadening determines the peak/dip position. Since, $R_D$ magnitude varies inversely with carrier density ($n^{-1.5}$), peak/dip appearing at higher densities with increasing temperature leads to allover slow variation with temperature rather than usual $T^2$ increase.

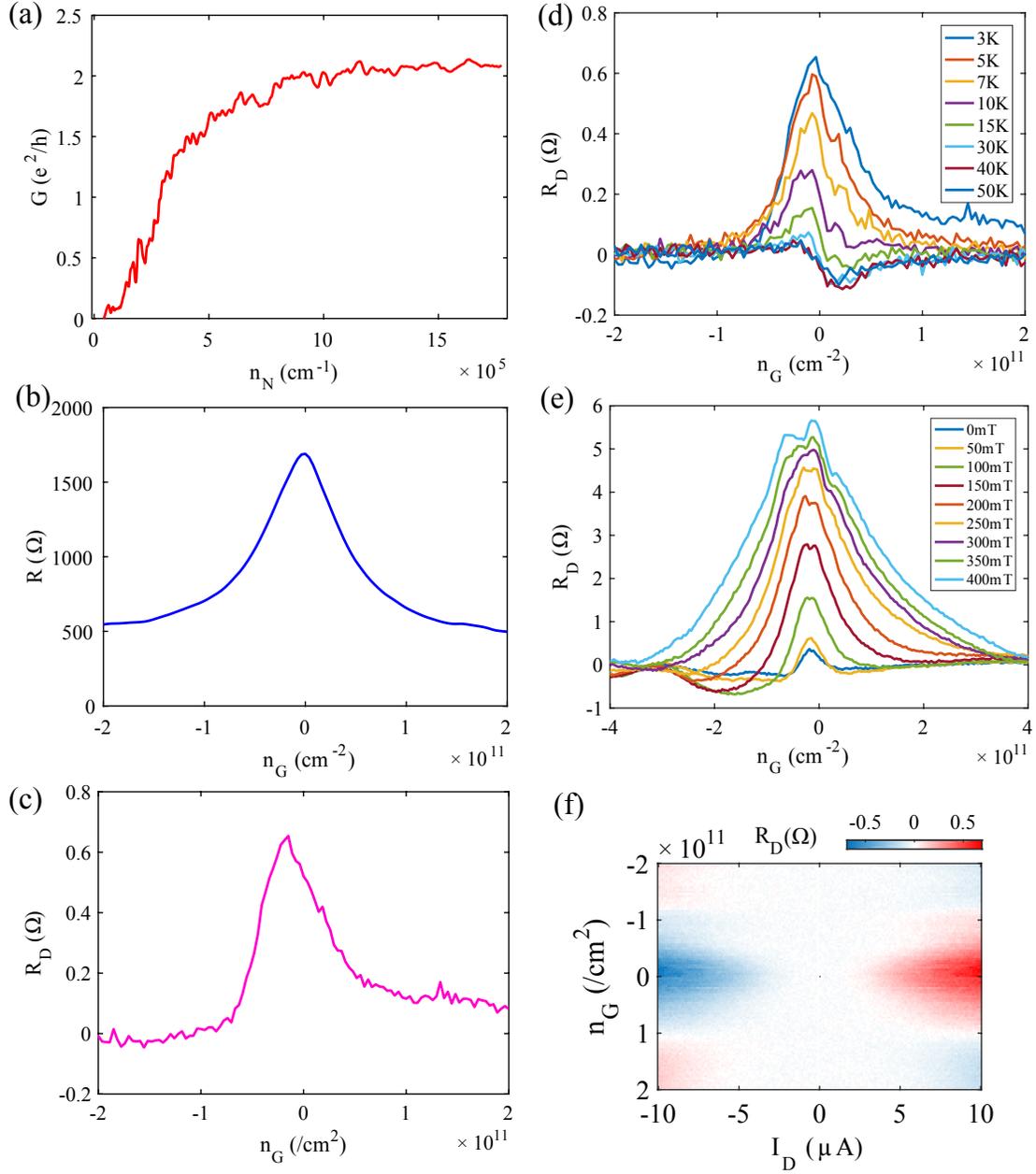

**Fig. S-4C: Data for MLG-NW (D1).** (a) NW conductance plot with the 1D nanowire density $n_N$ at T=1.5K. (b) and (c) respectively shows plot of 2-probe graphene resistance and Drag resistance $R_D$ plot with density $n_G$. (d) and (e) $R_D$ vs. $n_G$ plot for different temperatures and magnetic fields respectively. (f) 2D colormap of $R_D$ as a function of drive current $I_D$ and $n_G$. Data of (c)-(f) has been obtained for $n_N \sim 4 \times 10^5 cm^{-1}$.

20

## SI-4C: Density, Temperature and Magnetic field dependence of MLG-NW and BLG-NW devices

In this section we present nanowire and graphene response with density, magnetic field and temperature concomitantly with the drag resistance response. Fig. S-4C and 4D respectively shows the data where $R_D$ vs. $n_G$ plot is shown concomitantly with graphene and nanowire responses for MLG-NW D1 and BLG-NW D3 devices. Fig. S-4E has the data for nanowire conductance ((a), (b)) for different temperature and magnetic fields respectively.

## SI-5A: Calculation of $\dfrac{\partial Q}{\partial \mu}$ for Graphene and Nanowire

In this section we will discuss how we have obtained the expressions for $\dfrac{\partial Q}{\partial \mu}$ for the graphene and the NW and finally the total thermal conductivity of the graphene-NW system.

**Calculation of** $\dfrac{\partial Q}{\partial \mu}$**:** To explain the temperature dependence of our MLG-NW devices, a quantitative theory of ED in 2D-1D system is required. In the absence of such theory, we appeal to the Energy drag by Song et.al for 2D-2D systems. From ref [17], we obtain the expression for Energy drag as:

$$\rho_D \propto \frac{1}{2T\kappa}\left(\frac{\partial Q_{GR}}{\partial \mu_G}\right)\left(\frac{\partial Q_{NW}}{\partial \mu_N}\right) \tag{7}$$

Equation (7) shows that the drag resisivity is directly proportional to the $\dfrac{\partial Q}{\partial \mu}$ of both the graphene and NW layers and inversely related to the total thermal conductivity of both the layers $\kappa$. In the experiment, we observe the $R_D$ peak appearing at the Dirac point i.e. when $\mu_G = 0$, while the nanowire has a finite carrier density, i.e. $\mu_N \neq 0$. This situation indicates that in our devices only $\left.\dfrac{\partial Q_{GR}}{\partial \mu_G}\right|_{\mu_G=0}$ and $\left.\dfrac{\partial Q_{NW}}{\partial \mu_N}\right|_{\mu_N\neq 0}$ contributes towards the observed non-zero drag peak at the Dirac point.

In Energy drag the heat current ($\mathbf{j}_q$) and the charge current ($\mathbf{j}$) are coupled by: $\mathbf{j}_q = Q\mathbf{j}$, where $Q$ is the the Peltier coefficient. The general expression of $Q$ can be written in terms of the layer conductivity $\sigma$ and chemical potential $\mu$ as:

21

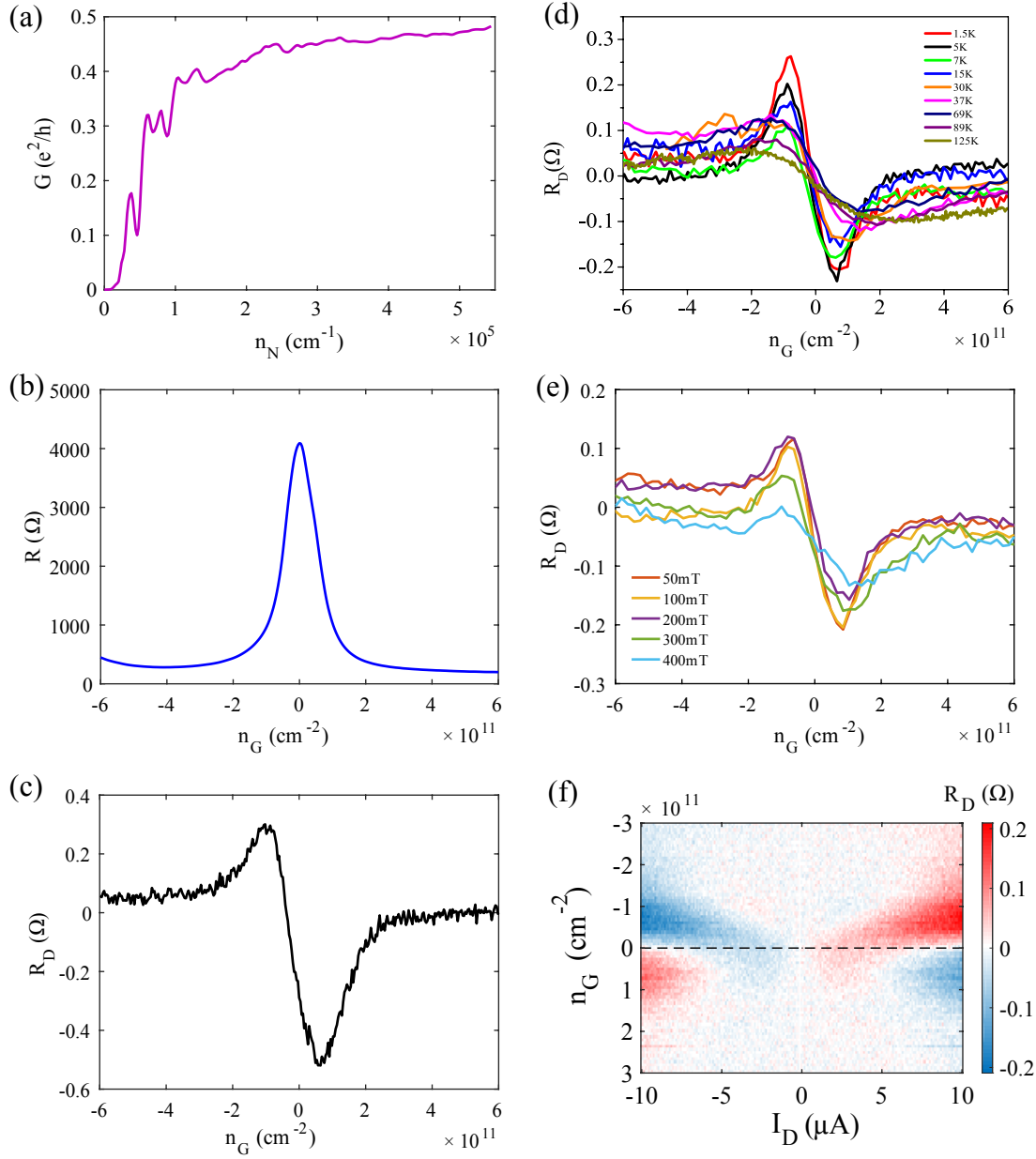

**Fig. S-4D: Data for BLG-NW (D3).** (a) NW conductance plot with the 1D nanowire density $n_N$ at T=1.5K. (b) and (c) respectively shows plot of 2-probe graphene resistance and Drag resistance $R_D$ plot with density $n_G$. (d) and (e) $R_D$ vs. $n_G$ plot for different temperatures and magnetic fields respectively. (f) 2D colormap of $R_D$ as a function of drive current $I_D$ and $n_G$. Data of (c)-(f) has been obtained for $n_N \sim 1 \times 10^5 cm^{-1}$.

$$Q = \frac{\pi^2 k_B^2 T^2}{3e} \frac{(d\sigma/d\mu)}{\sigma} \tag{8}$$

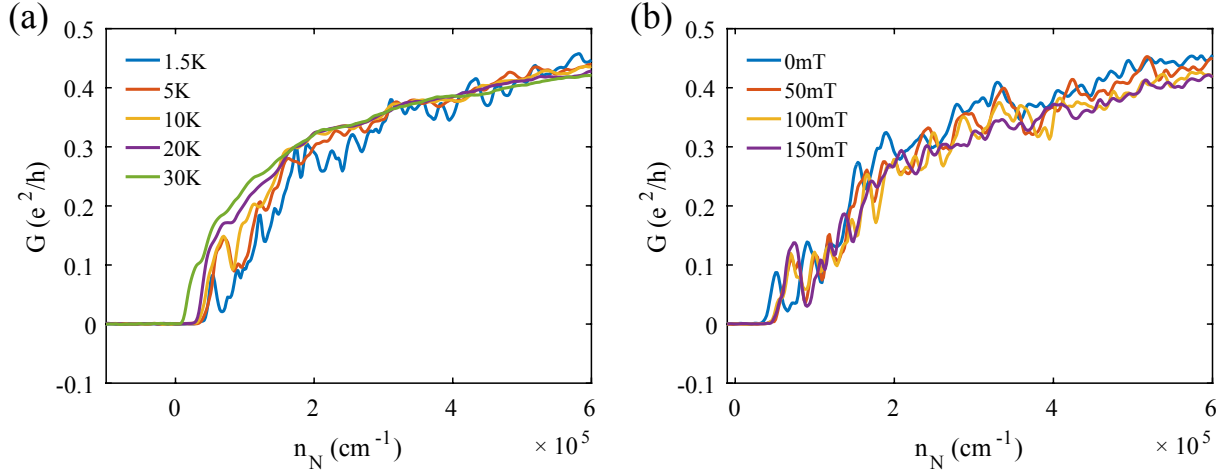

**Fig. S-4E** (a) and (b) NW conductance vs. $n_N$ plot for different temperatures and magnetic fields respectively. The NW belongs to device D2.

Now for graphene the conductivity can be expressed as $\sigma = \sigma_0 + n_G e \mu_{FE}$, where $\sigma_0$ is the residual conductivity at the charge neutrality point, $\mu_{FE}$ is the field-effect mobility and $n_G$ is the carrier density in graphene. Using this expression for conductivity and the linear energy-momentum relation for the monolayer graphene $\mu_G = \hbar v_f \sqrt{\pi n_G}$, equation (8) reduces to:

$$Q_{GR} = \frac{2\pi^2 k_B{}^2 T^2}{3e} \frac{\mu_G}{\mu_G^2 + \Delta^2} \qquad (9)$$

where $Q_{GR}$ is the Peltier coefficient for the monolayer graphene, $\Delta^2 = \frac{\sigma_0 \hbar^2 v_f^2 \pi}{e \mu_{FE}}$, here $\hbar$ is the reduced Plank constant, $v_f$ is the Fermi velocity, $k_B$ is the Boltzmann constant, $e$ being the electronic charge and $\mu_G$ is the graphene chemical potential respectively. The partial differentiation of equation (9) with respect to $\mu_G$ leads to $\left.\frac{\partial Q_{GR}}{\partial \mu_G}\right|_{\mu_G=0} \propto \frac{T^2}{\Delta^2}$. Putting appropriate parameters for graphene, i.e. $\sigma_0 \sim 400 \mu S$, $v_f = 10^6 m/s$, $\mu_{FE} = 100,000 cm^2/Vs$, we obtain $\Delta \sim 18 meV$.

Now to calculate $\frac{\partial Q_{NW}}{\partial \mu_N}$, we remember that InAs has parabolic band structure, i.e. $\mu_N = \frac{\hbar^2 \pi^2 n_N^2}{8m^*}$, where $m^*$ is the effective mass corresponding to the InAs band structure and we take the NW conductivity to be $\sigma = n_N e \mu_{FE}$ which yields:

23

$$Q_{NW} = \frac{\pi^2 k_B{}^2 T^2}{3e} \frac{1}{2\mu_N} \tag{10}$$

The partial differentiation of equation (10) leads to $\left.\frac{\partial Q_{NW}}{\partial \mu_N}\right|_{\mu_N \neq 0} \propto \frac{T^2}{\mu_N^2}$.

**Calculation of $\kappa$:** The total thermal conductivity $\kappa$ is the sum of the thermal conductivities of graphene ($\kappa_G$) and nanowire ($\kappa_N$), i.e. $\kappa = \kappa_G + \kappa_N$. Now for graphene, the electronic contribution towards $\kappa$ is dominant at low temperatures, so $\kappa_G \propto c_1 T$. Whereas for nanowires, the phononic contribution towards the thermal conductivity is very prominent, as the nanowires are poor electrical and hence poor thermal conductors. So, $\kappa_N \propto (c_2 T + bT^5)$, here $c1$, $c2$ and $b$ are constants. Combining the effects for both the drive and the drag layer yields $\kappa \propto (aT + bT^5)$, where $a = c_1 + c_2$ and $b$ are the relative contribution from the electronic and phononic part towards the thermal conductivity. This brings the equation (7) to:

$$\rho_D \propto \frac{T^3}{\mu_N^2 \Delta^2 (aT + bT^5)} \tag{11}$$

In this calculation we have not taken into account the contribution of the electron-phonon coupling from the interlayer dielectric hBN towards the total thermal conductivity $\kappa$ of the system. This is because the contribution from the mentioned effect will be much smaller than the corresponding contribution from nanowire and hence may not affect the temperature dependence of $R_D$.

Now to explain the temperature dependence of our MLG-NW devices from equation (11), we see that (shown in main text Fig 3(a)) the drag resistivity at first increases upto a certain $T'$ and monotonically decreases further where the $\frac{a}{b}$ ratio determines the value of $T'$. According to equation (11), the value of $T'$ is below our base temperature T=1.5K for $\frac{a}{b} = 5$ (shown in main text fig 3(a)), i.e. even when the phononic part is 5 times smaller then the electronic part.

### SI-5B. Plot of Q vs. $\mu$ and $\frac{\partial Q}{\partial \mu}$ vs. $\mu$ from experimental data

In the Fig. 3(b) of the main text, we have compared the density response of the $\frac{\partial Q_{GR}}{\partial \mu_G}$ derived from the experimental results to that of the measured drag resistance $R_D$ and we have found that the width of both

the plots near the Dirac point are very similar. This similarity in dependence supports our claim that the drag in MLG-NW devices are originated from the Energy drag mechanism [17]. In Fig. S-5 (a) the Peltier coefficient of the graphene $Q_{GR}$ vs. $\mu_G$ is plotted for one of the devices. We calculate $Q_{GR}$ from equation (8) from experimentally obtained gate response of the graphene. $\sigma$ vs. $\mu_G$ is obtained from the 2-probe resistance $R$ vs. $V_{GR}$ using $\sigma = \frac{L}{W}\frac{1}{R}$, where L and W are the length and width of the graphene channel and $\mu_G = \hbar v_f \sqrt{\pi n_G}$, $n_G$ being the graphene density; The conversion of $n_G$ from $V_{BG}$ has been discussed before in section SI-1C. Fig. S-5 (b) shows $\frac{\partial Q_{GR}}{\partial \mu_G}$ plot with $\mu_G$ which is calculated by performing derivative of equation (8):

$$\frac{\partial Q_{GR}}{\partial \mu_G} = \frac{\pi^2 k_B^2 T^2}{3e}\left[\frac{1}{\sigma}\frac{d^2\sigma}{d\mu_G^2} - \left(\frac{d\sigma/d\mu_G}{\sigma}\right)^2\right] \quad (12)$$

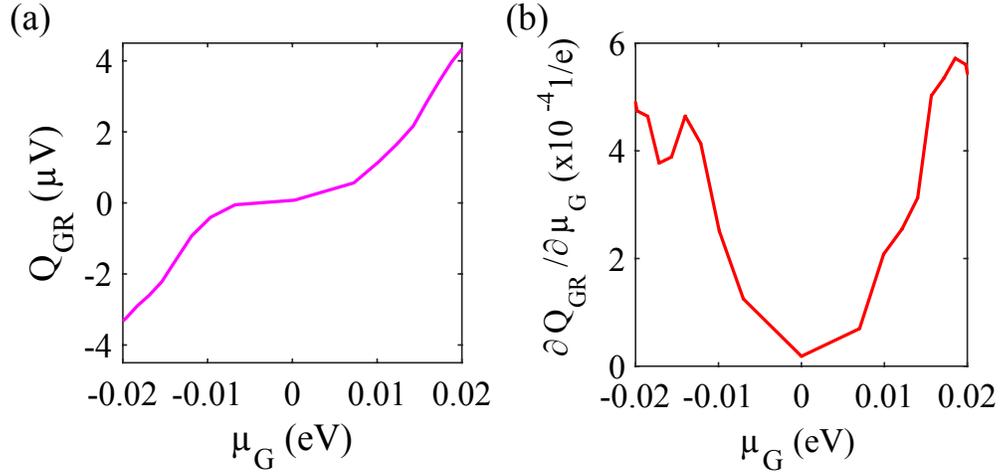

**Fig. S-5:** (a) Peltier coefficient ($Q_{GR}$) obtained from experimental data plotted with the chemical potential of graphene $\mu_G$. (b) Partial derivative of $Q_{GR}$ with respect to graphene chemical potential, $\partial Q_{GR}/\partial \mu_G$ plotted against $\mu_G$.

### SI-5C: Relation between the chemical potential ($\mu_N$) and carrier density ($n_N$) of the nanowire

In Fig. 1(g) of the main text, we see that the $R_D$ for MLG-NW devices varies with the nanowire density $n_N$ as $R_D \propto \frac{1}{n_N^4}$ and we say that this relation follows directly from the Energy drag mechnism in MLG-NW devices. To prove that, we recall equation (10), which says $Q_{NW} \propto \frac{1}{\mu_N}$. Partial differentiation of equation

25

(10) w.r.t $\mu_N$ yields $\left.\frac{\partial Q_{NW}}{\partial \mu_N}\right|_{\mu_N \neq 0} \propto \frac{1}{\mu_N^2}$. The relation between the chemical potential $\mu_N$ and the carrier density $n_N$ for InAs nanowire is $\mu_N \propto n_N^2$, which comes from the parabolic energy-momentum relation and the 1D density of states. Since, $\mu_N = \frac{\hbar^2 k_F^2}{2m^*}$ and $k_F = \frac{n_N \pi}{2}$ for one-dimensional system, where $k_F$ is the Fermi wavevector, $m^*$ is the effective mass of electron in the conduction band for InAs nanowire band structure, we obtain $\mu_N = \frac{\hbar^2 \pi^2 n_N^2}{8m^*}$. This leads to, $\left.\frac{\partial Q_{NW}}{\partial \mu_N}\right|_{\mu_N \neq 0} \propto \frac{1}{n_N^4}$ as obtained experimentally.


  					
\end{document}